\documentclass[11pt,a4paper]{article}

\usepackage[margin=2.5cm]{geometry}
\usepackage{amsmath,amssymb,amsthm,mathtools}
\usepackage{bm}
\usepackage[T1]{fontenc}
\usepackage{mathptmx,microtype}
\usepackage[authoryear]{natbib}
\bibpunct{(}{)}{;}{a}{,}{,}
\usepackage{booktabs,array,longtable,multirow,rotating}
\usepackage{enumitem}
\usepackage{xcolor}
\usepackage{tcolorbox}
\usepackage{algorithm,algpseudocode}
\usepackage{pdflscape}
\usepackage{graphicx}
\usepackage{float} 
\usepackage[hidelinks,
            pdftitle={DYSANOS: Generative Dynamic Smooth Arbitrage-free Non-parametric Option Surfaces},
            pdfauthor={Hans Buehler, Blanka Horvath, Anastasis Kratsios, and Magnus Wiese},
            pdfsubject={Quantitative Finance manuscript},
            pdfkeywords={option pricing, implied volatility, generative models, static arbitrage, risk-neutral measure, machine learning}]{hyperref}

\newtheorem{definition}{Definition}[section]
\newtheorem{theorem}[definition]{Theorem}

\newtheorem{remark}[definition]{Remark}

\newcommand{\R}{\mathbb{R}}
\newcommand{\N}{\mathbb{N}}
\newcommand{\E}{\mathbb{E}}
\newcommand{\Q}{\mathbb{Q}}

\newcommand{\calA}{\mathcal{A}}

\newcommand{\BSCall}{\mathrm{BSCall}}

\renewcommand{\P}{\mathbb{P}}
\renewcommand{\Q}{\mathbb{Q}}
\newcommand{\removeblock}[1]{}
\DeclareRobustCommand{\supstack}[2]{\stackrel{\mbox{{\scriptsize ${#1}$}}}{ {#2} }}     
\DeclareRobustCommand{\eqlabel}[1]{\label{eq:#1}}                                     
\DeclareRobustCommand{\eq}[1]{\begin{equation}\eqlabel{#1}}                           
\DeclareRobustCommand{\eqx}{\begin{equation}}                                         
\DeclareRobustCommand{\eqend}{\end{equation}}                                         

\title{DYSANOS: Generative Dynamic Smooth Arbitrage-free Non-parametric
Option Surfaces}
\author{%
  Hans Buehler\thanks{%
    Corresponding author. Mathematical Institute,
    University of Oxford. Email: \texttt{hans.buehler@maths.ox.ac.uk}}
  \and
  Blanka Horvath\thanks{%
    Mathematical Institute and Oxford-Man Institute,
    University of Oxford. \texttt{blanka.horvath@maths.ox.ac.uk}}
  \and
  Anastasis Kratsios\thanks{%
    McMaster University and Vector Institute.}
    \and
  Magnus Wiese\thanks{%
    Mathematical Institute,
    University of Oxford. \texttt{magnus.wiese@maths.ox.ac.uk}}
}
\date{}

\begin{document}
\maketitle


\begin{abstract}
This article presents with DYSANOS the first generative market model 
for smooth SANOS option surfaces for all strikes and expiries which are
free of static arbitrage. 
Our model is designed to generate entire paths of daily spot and option prices for years
in the future. By construction all its market shock scenarios are arbitrage-free.

We present a robust and useful if somewhat simplistic baseline in the form of an AR(1) model.
We discuss model setup, data pipeline, and training and investigate market
reconstruction, stability, and tail behavior. We illustrate model performance
on 891 Option Metrics IvyDB S\&P Index surfaces from 2022-01-03 through to
2025-08-29.

We also demonstrate how to use the methods of~\citet{removedrift} to construct
numerically a risk-neutral density. As part of this we develop a new test for zero conditional means under a given measure.
We show that for a set of 100,000 simulated paths
a trading universe of 48 options and spot is numerically free of dynamic arbitrage.
\end{abstract}

\noindent\textbf{Keywords:} option pricing; implied volatility; generative
models; static arbitrage; risk-neutral measure; machine learning

\medskip
\noindent\textbf{JEL classification:} C45; C58; G13

\bigskip


\section{Introduction}
\label{sec:intro}

The implied volatility surface of an option market changes dynamically over time, giving rise to
Vega risk when risk managing option portfolios.
Modelling the dynamics of a full continuous surface of options
in a way that is simultaneously \emph{statistically realistic} and \emph{statically
arbitrage-free} at every simulated date has remained an open problem since the
foundational work of~\citet{ContVol} on dynamics of the implied volatility surface.

\begin{figure}[H]
    \centering
    \includegraphics[width=0.9\linewidth]{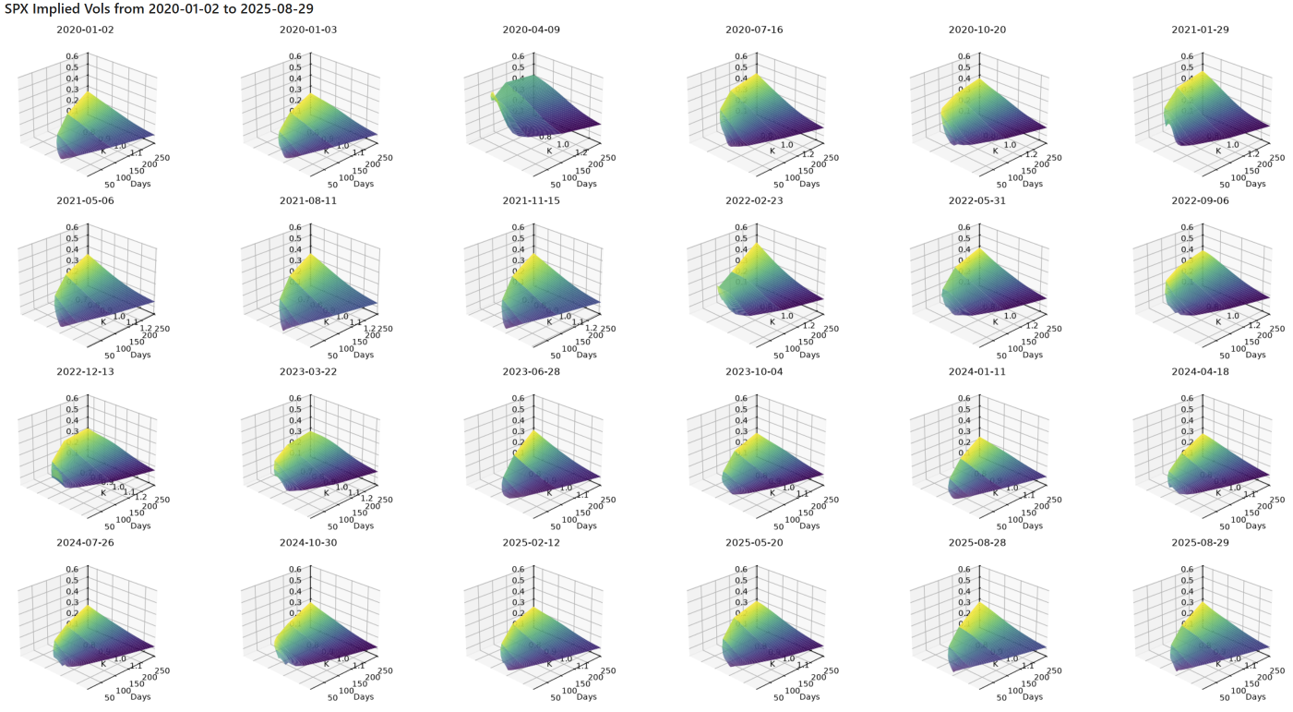}
    \caption{Historical implied volatilities for mid-prices of S\&P~500 Index options
    from 2020-01-02 to 2025-08-29
    using Option Metrics Ivy DB data via WRDS \url{https://wrds-www.wharton.upenn.edu/} (the latest data available at the time of writing).}
    \label{fig:initialexample}
\end{figure}

The most advanced approach until recently were the
simulators of~\citet{wiese2019simulate,wiese2021multiassetspotoptionmarket,Cohen02012023}
for trained generative dynamics of
a \emph{discrete floating grid} of options which are strictly
 arbitrage-free at each simulation date. These simulators are using, in essence, the discrete local volatility
representation pioneered in~\citet{DLV} to ensure absence of static arbitrage in every simulated
option surface. 

However, the restriction of the above models to a discrete set of floating option strikes and time-to-expiries
means that the simulators do not have access to the daily returns $dC_t(\tau,k)$
of actual tradable instruments with cash strikes $k$ and calendar expiries~$\tau$ (because the
contract terms change from one simulated day to another).  They therefore do
not directly provide statistically consistent time series of fixed-contract
returns, which are required to validate simulated dynamics against observed
market returns.
Since the discrete option prices are free of arbitrage, one method would be
to interpolate option prices linearly, which means all options would be free of static arbitrage.
However, as shown in~\citet{buehler2006expensive} and illustrated in~\citet{SANOS2026},
linear interpolation yields the most expensive possible arbitrage-free price for any
option off the grid, rendering the approach unsuitable in practice. Being linear
also means that the implied probability density is piecewise constant and zero between strikes.
In other words
the spot price can only take values at the model strikes for each expiry to avoid dynamic
arbitrage.

This article presents a model framework for
\begin{itemize}
    \item  \textbf{(i) realistic generative dynamics} of
    \item  \textbf{(ii) smooth option surfaces for all strikes and expiries} which are
    \item  \textbf{(iii) free of static arbitrage} and have
    \item  \textbf{(iv) a risk-neutral density}.
\end{itemize}

\subsection*{Historic Context and Literature Review}

\noindent
\textbf{The classical approach and its limitations:}
Quantitative finance has traditionally approached this problem by specifying
analytically tractable risk-neutral dynamics for the underlying asset ---
Heston, rough volatility, SABR --- and then \emph{calibrating} model parameters
to fit each day's option surface independently. They are then overlaid with 
a local volatility surface using the seminal construction of~\citet{Guyon2011TheSC}. The latter
guarantees that all options are initially priced correctly.

In this paradigm, realistic surface dynamics are an afterthought: calibration is repeated
from scratch on each date. 
This approach is both impractical for large-scale simulation and structurally inconsistent
with the goal of generating realistic implied volatility or option price
paths.\\

\noindent
\textbf{Simulating Implied Volatility:}
A natural data-driven alternative is to treat the implied volatility surface itself
as the simulation object, insipired first by~\citep{ContVol} and then implemented for example
by~\cite{ContVuletic2023,ChoudharyJaimungalBergeron2024,HanEtAl2026}.
The fundamental difficulty is that there is no known parametrization  of
the implied volatility surface which is strictly free of static
arbitrage. There exist numerous attempts to rectify this with classic approaches
(e.g.~\citealp{Kahale2004,fengler2009,GatheralJacquier2014,GuoJacquierMartiniNeufcourt2016,Orosi2015})
and machine-learned methods
(e.g.~\citealp{vae_ning2023arbitrage,AckererTagasovskaVatter2020,YangChenShuHospedales2025,GononJacquierWiedemann2025}),
but to our knowledge none of them can strictly guarantee absence of static arbitrage.

If a simulated surface admits a butterfly spread or calendar spread arbitrage,
then a learning agent can exploit an artificial profit which is unrelated to the
problem for which the generated market was intended. Aside from those classic constraints
call prices also need to adhere to boundary conditions which are often not discussed
in the literature (we repeat them below).

Most published works also simplify the problem by only considering few strikes and expiries.\\

\noindent
\textbf{Constrained Parameter Spaces and Code Books:}
The core challenge of any representation of an option price surface is that no-static-arbitrage conditions are essentially
linear conditions on call prices, which are difficult to enforce directly in a generative machine learning model. 
This challenge is shared by most representations including most parametrizations of implied volatility.

The idea of a `code book' as introduced in~\citet{TangentLevy} is to find a representation of the option price surface
which is numerically tractable for simulation and for which required parameter constraints are straightforward to enforce.
Indeed, the seminal construction of~\citet{TangentLevy} achieves that goal. In addition, the authors are able to also
construct a risk-neutral measure consistent with the code book representation.
They achieve this, essentially, by writing the spot price as a product of independent increments. That construction unfortunately also restricts
the dynamics their model can represent. It tends to fit poorly to statistical data due to the strong cross-expiry correlation of implied forward volatilities.
\\

\noindent
\textbf{Local Volatility as a Code Book:}
An intuitive alternative is to use Dupire's local volatility as a `code book' instead
\emph{(not as dynamical model)}. Given a non-negative local volatility function
$\sigma_t(T,K)$ for $(T,K)\in\R^2_{\geq 0}$ defined on some day~$t$
we can define 
a smooth, strictly arbitrage-free option surface 
$C_t(T,K):=C^\mathrm{LV}(\sigma_t;T,K)$ 
by solving the Dupire forward-PDE
$\partial_T C^\mathrm{LV}(\sigma_t;T,K)
 = \mbox{$\frac12$}K^2\sigma_t^2(T,K)\,
   \partial^2_{KK}C^\mathrm{LV}(\sigma_t;T,K)$.
That means any statistical generative model of Dupire's local volatility gives rise
to a dynamic model of option surfaces which are by construction free of static arbitrage.

However, in order to price any simulated option off a simulated local volatility surface
we would need to solve a forward-PDE on a suitably
fine grid which is slow, particularly when used within batch-based gradient descent training.

The idea to address this conundrum is to use a grid of options and define a discrete version
of local volatility on that grid, which is what we called
\emph{discrete local volatility (DLV)} in~\citet{DLV}.
It is essentially the discretized local volatility which is consistent with using an implicit forward-PDE operator on a sparse inhomogeneous grid of options,
as explained in~\citet{AndreasenHuge}.
However,
its native option prices for strikes not on its base grid are the result of linear interpolation and therefore, as mentioned above, `too expensive'.
We therefore use
our smooth SANOS extension~\citep{SANOS2026}.

The use of SANOS also allows to fit our model very efficiently to observed bid/ask spreads over the entire observed option surface.\\

\noindent
\textbf{Overview of Related Papers:}
Related work is summarized in Tables~\ref{tab:surface-interpolation-literature}
and~\ref{tab:surface-dynamics-literature}, starting on
page~\pageref{tab:surface-interpolation-literature}.

\subsection*{Key contributions}

\noindent
\textbf{SANOS Decoder:}
    We show how to implement and learn a coordinate transformation of
    our SANOS model~\citep{SANOS2026}
    such that the `ML-SANOS' state parameters $h_t$ live in an unconstrained space
    $\R^{d}$.  The call
    prices 
    $$
        h_t \longmapsto C(h_t;T,K)  \ .
    $$    
    remain smooth and statically arbitrage-free.
    
    We show how ML-SANOS can be trained directly to suitably pre-processed 
    option market data, yielding a trained $h_t$ for each day.
    The pre-processing step means that DYSANOS, via SANOS, is fitted to real market bid/ask prices.\\

\noindent
\textbf{Generative Dynamics:}
    We first show that the direct surface state can be reduced to an
    affine 200-dimensional representation with little loss of quote fit.  We
    then add a stable one-week AR(1) overlay and jointly refine the surface
    states and transition.  Its transition is parameterized as the matrix
    exponential of a dissipative generator, and its full innovation
    covariance is learned by a nonsingular Cholesky factor.  The overlay
    preserves the strong market fit while producing a stationary generative
    law.  That law permits exact Gaussian transitions at daily, one-week, or
    longer horizons; as non-Gaussian alternatives, we fit the Tail Diffusion,
    a no-replay scalar log-variance model, and resample historical one-week
    innovations.
    
    We do not
    claim that this approach provides the most realistic dynamics; rather,
    we see this implementation as baseline to beat with more modern
    time series models.\\

    \noindent
\textbf{Risk-Neutral Dynamics:}
Every decoded SANOS surface is free of static arbitrage by construction.
We show that we can `remove the drift' in the sense of~\citet{removedrift} for a subset of the (infinite) tradable instruments
by finding a martingale measure~$\Q$ equivalent to the learned~$\P$.
We test the resulting measure against two no-arbitrage criteria: zero
unconditional mean and zero conditional expected return. To assess the latter, we develop a novel `no-arb' test.
This is of independent interest.

The `remove the drift' procedure deployed here is brittle: it is sensitive to numerical instabilities,
errors, overfitting and other numerical challenges. While we did manage to find a risk-neutral
measure for the paths we generate, it was an iterative process. As it is the method is not suitable
for a production environment.

\section{SANOS}

\label{sec:markets}

We consider European call options on a single underlying
for historic days $t\in \{t_1,\ldots,t_m\}\subset \N$
which we will count in business days.
We assume forwards, discount factors, and survival probabilities
are deterministic. As described in remark~\ref{rem:driftless},
 we can therefore convert them at each point~$t$ to driftless equivalents
where spot, forwards, discount factors and survival probabilities are all one, for every~$t$.

At time~$t$, we observe~$L_t\in\N$ 
European call option prices with time-to-expiries $T_t^\ell \geq 0$ and
relative strikes $K_t^\ell\geq 0$
for $\ell=1,\ldots,L_t$
in the market. Their symbolic payoff is therefore
$$
   \left(\frac{S_{t+T_t^\ell}}{S_t} - K_t^\ell \right)^+ \ .
$$
We denote by $B_t^{\ell}$ and $A_t^\ell$ their bid and ask prices;
mid-price and half-spread are defined (in vector notation) as
$C_t := \tfrac{1}{2}(A_t + B_t)$ and
$\gamma_t := \tfrac{1}{2}(A_t - B_t)$.

We generally assume that $B_t\succ0$ and $\gamma_t\succ0$.\footnote{
We define $x \succ y \Leftrightarrow \forall i: x^i > y^i$.}
Contrary to most derivative literature, we do not assume that bid/ask prices are in some way free of arbitrage; instead
we will show how to handle arbitrage valuations gracefully.

We will use the Black~\&~Scholes formula in terms of \emph{total volatilities} defined 
for a spot $S$, strike~$K$ and total volatility~$W$ as
\begin{equation}\label{eq:bs}
    \BSCall(S,K,W) := S\,\mathcal{N}( d_+ ) - K\,\mathcal{N}(d_-)
    \ \ \ \mbox{for} \ \ \
    d_\pm := \frac{ 1 }{ W } \log \frac S K \pm \frac12 W
\end{equation}
where~$\mathcal{N}$ denotes the normal distribution function.

\begin{remark}[Conversion to driftless data]\label{rem:driftless}
  Assume that at each time $t$ spot is $\tilde S_t$, forwards for some time to expiry~$T\geq 0$ are $\tilde F_t(T)$, discount factors are $\tilde B_t(T)$.
  In  that case the spot at $t+T$ as seen from time $t$ must be given as
  $$
  \tilde S_T^{(t)}= \tilde F_t(T)X_T^{(t)} 
  $$
  for a non-negative martingale $X^{(t)}$ with $X^{(t)}_0=1$. 

  Assume then that $\tilde C_t( T,\tilde{K})$ denotes the market call price for time-to-expiry~$T$ and cash strike~$\tilde K$. 
  Define the relative strike as
  $
  K :=\tilde K / \tilde F_t(  T) 
  $.

  The driftless call price is then given as
  $$
    C_t( T,K) := \frac{\tilde{C}_t\Big( T,\, K \tilde F_t(T) \Big)}{\tilde B_t(T) \tilde F_t(T) } \ .
  $$ 
  Puts are defined equivalently.
  This results can  be extended to the presence of cash dividends and deterministic survival probabilities.\footnote{
    If cash dividends 
    are present and if its survival probabilities are deterministic functions $\tilde Q_t(T)\equiv \exp( - \lambda_t(T)T )\in [0,1]$ then it was shown in~\citet{BuehlerVolAndDivs}
    that spot at $t+T$ as seen from time $t$ must be given as $\tilde S_T^{(t)}=\big(\tilde F_t(T)-\tilde D_t(T)\big) X_T^{(t)} + \tilde D_t(T)$, conditional on survival.
    The deterministic function $\tilde D_t(T)$ represents floor enforced on the spot price by the presence of future cash dividends, cf.~\citet{BuehlerVolAndDivs}.
    for specifics.

    The driftless strike now becomes 
    $
    K := (\tilde K - \tilde D_t( T) ) / ( \tilde F_t(  T) - \tilde D_t(T) )
    $
    for which the driftless call price is now
    $$
      C_t( T,K) := \frac{\tilde{C}_t\Big( T,\, K\big(\tilde F_t(T)-\tilde D_t(T)\big) + \tilde D_t(T)\Big)}{\tilde B_t(T) \, \tilde Q_t(T)\,\big(\tilde F_t(T)-\tilde D_t(T)\big)} \ .
    $$
    For puts a similar formula applies after applying put-call parity; see~\citet{BuehlerVolAndDivs}.}
\end{remark}

\begin{remark}[Conversion to driftless spot returns]\label{rem:driftlessspot}
  A consistent driftless spot return series can be obtained by the fully funded one-day forward-return
  \eq{driftlessspotret}
    \frac{ S_{t+1} }{ S_t } := \frac{\tilde S_{t+1}}{\tilde F_t(t+1)}  \ .
  \eqend
\end{remark}

\subsection{Arbitrage-Free Call Price Functions}

A~call price function is simply a function $C:\R^2_{\geq} \longrightarrow  \R_{\geq 0}$
which computes call prices $C(T,K)$ for time-to-expiries $T\geq 0$ and relative
strikes $K\geq 0$ (i.e.~the call prices are normalized to a spot price of~$1$). 
The objective of this section and the next is to generate such functions $C_t$ for each
simulated date~$t$ such that each $C_t$ is `free of static arbitrage' in the following sense:

\begin{definition}
A call price surface $C: \R^2_{\geq 0}\to \R_{\geq 0}$ is \emph{free of static arbitrage} if there is no trade
which has non-negative returns with a non-zero probability of
positive returns.    

This is equivalent to the existence of a
martingale~$Z$ such that $C(T,K)= \E\big[(Z_{T}-K)^+\big]$.
\end{definition}

\begin{theorem}\label{th:arbfree}
A call price surface $C$ is free of static arbitrage if and only if 
it does not admit negative butterfly or calendar
spreads and satisfies the boundary conditions
$C(T,0)=1$, $C(0,K)=(1-K)^+$, $\lim_{K\uparrow\infty} C(T,K)=0$ and $\partial_K C(T,0)\in[-1,0)$.

The stricter condition~$\partial_K C(T,0)=-1$ applies if the stock price cannot become zero.
\end{theorem}

This has been proven in~\citet{buehler2006expensive}. Remark~\ref{rem:actionable_arbitrage}
shows that these conditions each are not superfluous: we give a practical
guide on how to construct \emph{actionable} arbitrage strategies if any of the above
are violated. 

\begin{remark}[Actionable arbitrage if the conditions of theorem~\ref{th:arbfree} are not met]\label{rem:actionable_arbitrage}
All of the below are meant as practical heuristics.
\begin{itemize}
    \item \textbf{Butterfly arbitrage:} this means that the price of a positive
    butterfly payoff is negative. The arbitrage is to get paid to enter into 
    the butterfly which can only ever generate a positive return.

    \item \textbf{Calendar arbitrage:} if for the same strike $K$ 
    the call price at $T_2$ is below the call price
    at $T_1$, then:
    \begin{enumerate}
        \item 
            Sell the call at $T_1$ and buy the call at $T_2$. This provides
            a positive cash flow $\pi_0>0$.
        \item 
            In $T_1$ if the call expires out of the money, then we have made money: both
            $\pi_0$ and a potential positive payoff at $T_2$ in our favour.

            If the call expires in the money, then we deliver one share
            and receive $K$. We are now short that share.
            In $T_2$, if the call expires in the money, then we receive one share
            to cover our short in return for $K$. Our total cashflow is $\pi_0$.
            If the call expires out of the money, then $S_T<K$. We can therefore
            buy back the share and cover our short for less than $K$.
            Our profits are $\pi_0+(K-S_T)^+>0$.
    \end{enumerate}
    \item If $C(0,K) \not= (1-K)^+$ we have an immediate pathological arbitrage.
    \item If $C(T,0)>1$ then sell the call and cover with one share; if $C(T,0)<1$
    then buy the call and enter into a short equity position.

    \item $\partial_K C(T,0)\geq -1$: assume on the contrary there is a $K\downarrow 0$
    such that $(1-C(T,K))/(-K) < -1$ i.e.~$C(T,K) = 1-K - \epsilon$ for some $\epsilon>0$.
          Buy the call and short sell the equity, resulting a cash position of $\pi_0=-K+\epsilon$;
          the payoff in $T$ is $(S_T-K)^+$ which allows us to cover our position and cash
          with a total reward of~$\epsilon>0$.

    \item $\partial_K C(T,0)<0$: assume on the contrary there is a $K\downarrow 0$
    such that $(1-C(T,K))/(-K) \geq 0$. We show existence of
    arbitrage in the case of equality
    i.e.~existence of some~$K>0$ such that $C(T,K)=1$.
    In that case we sell the call and buy the equity for~$1$.
    If the call expires in the money, we deliver the share
    and receive $K>0$. If the call expires out of the money, then sell
    the share for $S_T\geq 0$.

    \item $\lim_{K\uparrow \infty} C(T,K) = 0$ which means for all $c>0$
    there exists some $K$ such that $C(T,K')<c$ for all $K'\geq K$ (because of convexity).
    This is the only arbitrage condition which is technical because a violation means, 
    essentially, that the equity can grow without bounds which violates
    integrability.
    Assume therefore on the contrary 
    that there is some~$c>0$ such that $C(T,K)\geq c$ for all $K$. 
    Notice that, formally,
    $$
        S_T = - \int_0^\infty \frac{ (S_T-K+dK)^+ - (S_T-K)^+ }{ dK } dK
            = S_T + \lim_{K\uparrow \infty} (S_T-K)^+ \ .
    $$
    Therefore sell (a discretization of) the strip
    $$
    \int_0^\infty \frac{ C(T,K) -  C(T,K+dK) }{ dK } dK = 1 + e
    $$
    which delivers~$S_T$, and buy the equity. This will generate profits of~$e>0$.
 \end{itemize}
\end{remark}

\subsection{Smooth Arbitrage-free Non-parametric Option Surfaces}
\label{sec:sanos}

In~\citet{SANOS2026} we introduced a non-parametric family
of call price functions which are all statically arbitrage-free. For clarity, let us briefly fix $t$ and drop it from the notation of market data
and model parameters.

Each call price function is given in terms of model expiries
$0 = \hat T^0 < \hat T^1 < \cdots < \hat T^M$ and, for each model expiry,
$N^j>0$ inner strikes.\footnote{Note that we changed notation slightly from~\citet{SANOS2026}.}  Including the two boundary strikes, write
$\hat K^j = (\hat K^{j,i})_{i=0,\ldots,N^j+1}$. It is recommended that the two outer strikes
are far away from the inner strikes.\footnote{
    In our experiments, the inner strikes
    $0<\hat K^{j,1}<\cdots<1<\cdots<\hat K^{j,N^j}$
    were chosen first and then extended by $\hat K^{j,0}:=\hat K^{j,1}/2$ and
    $\hat K^{j,N^j+1} = 1.5\,\hat K^{j,N^j}$.
}  Model strikes and expiries do not need to match
market strikes or expiries. See~\citet{SANOS2026} for general comments on grid construction.

We say that $q=(q^1,\ldots,q^M)$ with $q^j\in\R^{N^j+2}$ is a \emph{martingale density} if
it satisfies the linear constraints
\begin{equation}\label{eq:qocnd}
    \left\{
    \begin{array}{ll}
        \mbox{Each $q^j$ is a density:} & q^j\geq 0,\ (q^j)'1 = 1 \\
        \mbox{Each $q^j$ has unit expectation:} & (q^j)' \hat K^j=1 \\
        \mbox{The $q$ are in convex order:}&
            {q^j}' CX_j \geq {q^{j-1}}' CX_{j|j-1}
            \\
    \end{array}
    \right.
\end{equation}
The matrices $CX_j$ and $CX_{j|j-1}$ project the density on call prices. They are defined as
$$
CX_j^{i,k} := (\hat K^{j,i}-\hat K^{j,k})^+ \ \ \ i=0,\ldots,N^j+1;\ k=1,\ldots,N^j 
$$
and
$$
CX_{j|j-1}^{i,k} := (\hat K^{j-1,i}-\hat K^{j,k})^+ \ \ \ i=0,\ldots,N^{j-1}+1;\ k=1,\ldots,N^j  \ . 
$$
Accordingly, ${q^j}' CX_j \in \R^{N_j}$ is the vector of call prices for strikes $\hat K^{j}\in\R^{N_j}$ computed using the density $q^j$
while ${q^{j-1}}' CX_{j|j-1} \in \R^{N_j}$ is the vector of call prices for the same strikes $\hat K^{j}\in\R^{N_j}$ computed using the previous density $q^{j-1}$.
The condition ${q^j}' CX_j \geq {q^{j-1}}' CX_{j|j-1}$ enforces that call prices are increasing into each strike $\hat K^{j}$.

\begin{definition}[SANOS call price]
\label{def:sanos}
Let $q$ be a martingale density and $0<W^1< \cdots < W^M$
be increasing smoothing total volatilities.\footnote{The linear-programming preprocessing uses 50\% of ATM total volatility. The trained ML-SANOS decoder instead learns the positive forward smoothing multipliers $\eta^j$.}
The SANOS call price 
 for arbitrary $T\geq0 ,K\geq 0$ 
is given by
\begin{equation}
  \label{eq:sanos_interp}
  \begin{aligned}
  C(T,K):={}&\alpha^j(T)\sum_{i=0}^{N^j+1}
       \BSCall\!\bigl(\hat K^{j,i},K,W(T)\bigr)q^{j,i}\\
     &+(1-\alpha^j(T))\sum_{i=0}^{N^{j-1}+1}
       \BSCall\!\bigl(\hat K^{j-1,i},K,W(T)\bigr)q^{j-1,i},
       \qquad T \in (\hat T^{j-1},\hat T^j].
  \end{aligned}
\end{equation}
with $\alpha^j(T) = (T - \hat T^{j-1})/(\hat T^j - \hat T^{j-1})$
and total volatility $W(T) := \alpha^j(T) W^j + (1-\alpha^j(T)) W^{j-1}$.

For $T>\hat T^M$ we
set $\alpha^{M+1}(T)=0$ and $W(T):=W^M\sqrt{T / \hat T^M}$; however extrapolation beyond
the last model expiry or before the front expiry is generally not recommended.
\end{definition}

\begin{theorem}
    The call price function $C$ is statically arbitrage-free.
For $W(T)>0$ it is $C^\infty$ in $K$. The function is as smooth as~$\alpha$ in~$T$.
\end{theorem}

\begin{remark}
    Contrary to~\citet{SANOS2026}, we interpolate the Black \& Scholes backbone in total volatility linearly, not variance.
    This is a mechanical choice to avoid taking square roots of variables during training
    which can lead to exploding gradients. If forward volatilities are bounded well away from
    zero, linear interpolation in variance is suitable, too.
\end{remark}

\begin{remark}
    The Black~\&~Scholes formula in~\eqref{eq:sanos_interp} can be replaced
    by the call price function for any martingale in~$T$, for example the FFT Heston price formula~\citep{carrmadan1999}.
    For practical implementations, the
    function should be fast on a GPU.
\end{remark}

\paragraph{Calibration as a linear program:}
 We assume we obtain $W^1<\cdots<W^M$
from the market e.g.~as square roots of ATM variance per expiry.
Given~$W$ the model call price $C^\ell := C(T^\ell, K^\ell)$ for a market expiry $T^\ell$ and
a market strike $K^\ell$
can be seen in~\eqref{eq:sanos_interp}
to be linear in~$q$.
The market-fitting problem can therefore be formulated as the linear programme
\begin{equation}
  \label{eq:lp}
  \min_{q} \sum_\ell \frac{
    (C^\ell - \bm{A}^\ell)^+ + (\bm{B}^\ell - C^\ell)^+ + \epsilon |C^\ell - \bm{M}^\ell|
  }{ (A^\ell - B^\ell)/2}
\end{equation}
subject to the linear constraints~\eqref{eq:qocnd} encoding non-negativity, normalisation, integration to~$1$, and the inter-expiry convex-order constraints that enforce absence of calendar-spread
arbitrage. We note that the model strikes and time-to-expiries do not need to correspond
to market strikes or expiries.\\

As discussed in~\citet{SANOS2026}, the primary application of this model is \emph{interpolation}: if the model strikes
are a superset of \emph{all} strikes of relevant market instruments then the model can practically be fit to the
vast majority of options across the entire surface from 1 day to expiries to several years. Stronger, if the model is
linear ($W=0$) the existence of the densities $q$ is equivalent to absence of arbitrage with respect to the market bid/asks (usually
this simply means our market data is not aligned correctly).

\begin{figure}[H]
    \centering
    \includegraphics[width=0.9\linewidth]{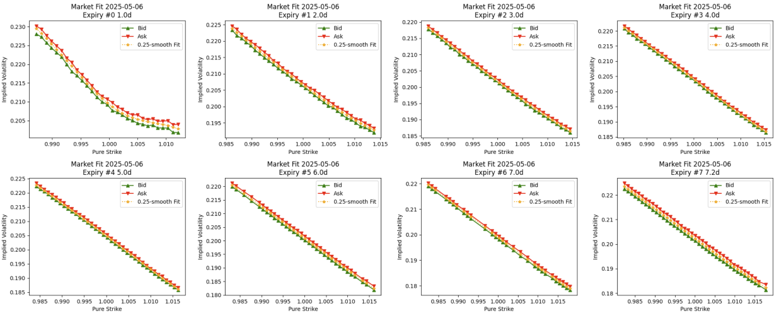}\\
    \includegraphics[width=0.9\linewidth]{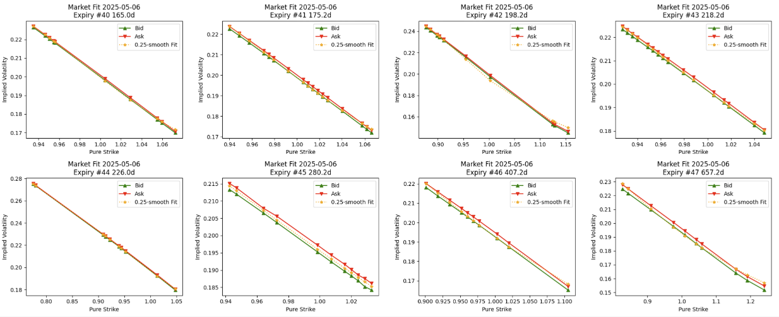}
    \caption{Example fit of SANOS.}
    \label{fig:sanos_fit}
\end{figure}

On S\&P~500 Index data with 1000 options across 48 expiries (2-day to 657 business-day TTE),
the LP solves in sub-seconds on a desktop PC, achieving 91.4\% of options within
bid--ask on a representative date (2025-05-06), with a median error of 21\% of the
half-spread on the remaining options. Figure~\ref{fig:sanos_fit} illustrates
the quality of the fit. The example is from~\citet{SANOS2026}; refer to that work for further details.

\subsection{ML-SANOS}
\label{sec:mlsanos}

We may view our  SANOS call price operator as a map from the model
parameters to a function space
\begin{equation}\label{eq:sanosmap}
    \big( q, W;\, \hat K,\hat T\big) \supstack{\mathrm{SANOS}}\longmapsto C    
\end{equation}
for which we naturally want to learn $q,W$ and, as we will see, $\hat K$ (as the right strike grid
depends on the implied variance of the market).
However, the density weights $q^j$ in particular are subject to the linear constraints~\eqref{eq:qocnd}
for which no exact methods exist to integrate them into standard batch gradient descent methods.
In this section we therefore reparameterize the map~\eqref{eq:sanosmap} in terms
of learnable parameters.

To do so, we no longer require strict \emph{interpolation} of the market prices. While we will want to be as close as possible
to observed market prices (and will show that we are able to do so), we now focus on a model parametrization which lends itself
to efficient learning using standard optimization techniques. That means that the number of strikes must somehow be kept manageable,
rather than be dictated by the number of market instruments at each observed historic day~$t$.

As a first step we therefore fix a common strike grid $\hat K$ with $0<\hat K^0<\cdots<1<\cdots<\hat K^{N+1}$ across expiries to simplify our model.
The call price for strike $\hat K^i$ given the density $q^j$ is given as $c^{j,i} = (q^j)'(\hat K - \hat K^i)^+$.
The convex order condition in~\eqref{eq:qocnd} therefore becomes simply $c^{j,i} \geq c^{j-1,i}$ for all $i$.

We now reparameterize our densities in terms of \emph{discrete local volatilities (DLVs)}~\citep{DLV}: let
$$     
    \omega^{i\pm} := \pm    \frac{ (\hat K^{i})^2 }{ (\hat K^{i+1} - \hat K^{i-1})\,(\hat K^{i\pm 1} - \hat K^{i}) }  \geq 0
$$
and with it the tridiagonal matrix~$\Omega$ with 
\begin{itemize}
    \item Upper diagonal $( -\omega^{1-}, \ldots, -\omega^{N-}, 0 )$,
    \item Lower diagonal $( 0, -\omega^{1+}, \ldots, -\omega^{N+} )$, and
    \item Main diagonal $( 0, +\omega^{1+}+\omega^{1-}, \ldots, +\omega^{N+}+\omega^{N-}, 0 )$.
\end{itemize}
i.e.
\eqx
    \Omega = \left(
                \begin{array}{cccccccccc}
                0 & -\omega^{1-} & 0 \\
                0 & +\omega^{1+}+\omega^{1-} & -\omega^{2-} \\
                0 & -\omega^{1+} & +\omega^{2+}+\omega^{2-} \\
                & & & &  \ddots\\
                & & & &  \ddots\\
                & & & & & +\omega^{(N-1)+}+\omega^{(N-1)-} & -\omega^{N-} & 0 \\
                & & & & & -\omega^{(N-1)+} & +\omega^{N+}+\omega^{N-} & 0 \\
                & & & & & 0 & -\omega^{N+} & 
                \end{array} 
    \right) \ .
\eqend
Then for \emph{any} squared (discrete) local volatilities $\Sigma^j = ( 0, \Sigma^{j,1}, \ldots, \Sigma^{j,N}, 0 )$
the operator
\eqx
            1 + \Omega\ \mathrm{diag}(\Sigma^j) (\hat T_j - \hat T_{j-1}) \in \R^{(N+2)\times(N+2)}
\eqend
is a tridiagonal operator with positive main diagonal and negative off diagonals. 
It is basically the discrete implicit scheme for the forward equation of the continuous time Dupire SDE.
It is well known~\citep{DLV}
that this operator is invertible, and the inverse~$\Xi^j$ can be efficiently computed on GPU using a compiled
Thomas' algorithm. It is also
a martingale transition operator in
 the sense of~\citet{DLV}, which means that the densities $q^1,\ldots,q^M$ generated
 iteratively starting in $q^0=(1)$ by
\begin{equation}\label{eq:densitytrans}
    q^j = \Big( 1+  \Omega\ \mathrm{diag}(\Sigma^j) (\hat T_j - \hat T_{j-1}) \Big)^{-1} q^{j-1} 
\end{equation}
form a martingale density over~$\hat K$.

With the above we have established a map from much simpler non-negative discrete local vols squared to
densities, and then to SANOS call prices:
\begin{equation}\label{eq:sanosmap2}
    \big( \Sigma, W;\, \hat K,\hat T\big) \supstack{\mathrm{DLV}}\longmapsto
    \big( q, W;\, \hat K,\hat T\big) \supstack{\mathrm{SANOS}}\longmapsto C \ .
\end{equation}
That allows us to pass to a practically unconstrained parameter set, which we call the ML-SANOS parametrization.
We will use the sigmoid function $\mathrm{sigmoid}(x)=(1+e^{-x})^{-1}$.

\begin{definition}[ML-SANOS]\label{def:ml-sanos}
    Fix the normalized grid geometry for ML-SANOS with:
    \begin{itemize}
        \item Time-to-expiries (TTEs) $0=\hat T^0<\cdots<\hat T^M$ with
        $\Delta\hat T^j = \hat T^j-\hat T^{j-1}$;
        \item Normalized strikes $z^1<\cdots<0<\cdots<z^N$ which we assume contain zero;
        \item Minimum and maximum forward volatilities $\sigma_{\max} > \sigma_{\min} > 0$; 
        \item Minimum and maximum squared discrete local volatilities $\Sigma_{\max} > \Sigma_{\min} > 0$; and
        \item Minimum and maximum smoothing factors $1> \eta_{\max} > \eta_{\min} \geq 0$.
    \end{itemize}
    We translate the new real-valued \emph{ML-SANOS parameters}
    \eq{production-native-state}
      u=(u^\Sigma,u^\sigma,u^{\eta})    \in\R^{M\times N+M+2}
    \eqend
    into SANOS parameters as follows:
    \begin{itemize}
        \item Total volatilities: 
        define the \textbf{forward volatilities}
             \eqx
                    \sigma^j := \sigma_{\min}
               +\mathrm{sigmoid}( u^{\sigma,j} ) ( \sigma_{\max} - \sigma_{\min} )
            \eqend
            and two smoothing factors
            \begin{equation}
                \label{eq:etatrans}
                \eta^{\mathrm f}:=\eta_{\min}
                  +(\eta_{\max}-\eta_{\min})
                    \mathrm{sigmoid}(u^{\eta,1}) \ \ \ \mbox{and} \ \ \
                \eta^{\mathrm{term}}:=\eta_{\min}
                  +(\eta_{\max}-\eta_{\min})
                    \mathrm{sigmoid}(u^{\eta,2}) \ .
            \end{equation}
            With these quantities define the \textbf{total volatilities}
             \begin{equation}\label{eq:totalW}
             \widetilde W^j := \sum_{\ell=1}^{j} \sigma^\ell \sqrt{\Delta\hat T^\ell}
             \qquad\mbox{and}\qquad
             W^j := \eta^{\mathrm f} \sigma^1 \sqrt{\Delta\hat T^1}\ + \eta^{\mathrm{term}} \sum_{\ell=2}^{j}  \sigma^\ell \sqrt{\Delta\hat T^\ell}\ .
             \end{equation}
        \item We use $\widetilde W^j$ to define a fixed grid as \textbf{relative strikes}
              \begin{equation}\label{eq:constructK}
                \hat K := \exp\left( z \widetilde W^1 \right) \bigcup \exp\left( z^i \widetilde W^M \right) \ ,
              \end{equation}
              sufficiently sorted, 
              with boundaries $\hat K^{0}:=\frac12 \hat K^{1}$ and $\hat K^{N+1} := \frac32 \hat K^{N}$.
        \item \textbf{Squared discrete local volatilities}:
              \begin{equation}\label{eq:defDLVdsq}
                 \Sigma^{j,i}
                 :=\min\!\left\{\Sigma_{\max},
                    \Sigma_{\min}+\exp(u^{\Sigma,j,i})\right\}\ .
              \end{equation}
              (This formulation avoids saturation of the sigmoid function as the raw log-DLV coordinate approaches the cap  $\Sigma_{\max}$;
              it requires adding a penalty term to the loss function.)

    \end{itemize}
\end{definition}
In our
experiments we use the bounds
$(\Sigma_{\min},\Sigma_{\max})=(0.01^2,3^2)$,
$(\sigma_{\min},\sigma_{\max})=(0.01,1)$, and
$(\eta_{\min},\eta_{\max})=(0.01,0.50)$. 

Put together, this process describes a map
    \begin{equation}
        \big(\,u;z,\hat T\,\big)  \supstack{
            \begin{array}{c}
                \mathrm{ML}\\
                \mathrm{SANOS}
                \end{array}
        }\longmapsto \big(\, \Sigma,W;\,\hat K,\hat T \big) 
        \supstack{\mathrm{DLV}}\longmapsto \big(\,q,W;\,\hat K,\hat T \big)  \supstack{\mathrm{SANOS}}\longmapsto C \ .
    \end{equation}
    In our new parametrization $u$ lives in an unrestricted space and is therefore
    particularly suited for ML/AI based learning.  The expiry locations remain fixed.

\section{DYSANOS}
\label{sec:data}

Given market data for $t=t_1,\ldots,t_m$ we now aim to estimate a dynamic model
for the parameters~\eqref{eq:production-native-state} for each day,
\begin{equation}
  u_t=\left(
u_t^{\Sigma},\,
            u_t^{\sigma},\,
      u_t^{\eta}
      \right)
  \in\R^{M(N+1)+2}.
\end{equation}
We use $M=18$ model expiries given by
\begin{align*}
    (\ 1,2,3,4,5,10,15,\ldots,70\ ) 
\end{align*}
business days.
We use $N=200$ strikes computed using~\eqref{eq:constructK} for $z$ strikes within $(-3,2)$.
The model therefore has 3,620 native
coordinates.  \\

\noindent
\textbf{Data and model geometries:}
The empirical population sourced from WRDS' Option Metrics Ivy data base contains 891 S\&P~Index dates from 2022-01-03 through
2025-08-29. See appendix~\ref{app:data-pipeline} for our baseline pipeline we use for all our research with this data set.
In addition to the filters described there we here also exclude options outside the model's strike range (which is effectively $(-3,2)$ NMS) and any options expiring
past the last model expiry of~70 business days. We also exclude tomorrow's AM-settled options. 

We obtain 3,566,258 original call-or-put book observations for our model.
All options are represented in call-equivalent units through put–call parity; call and put books at the
same expiry and strike remain separate observations and contribute separate loss terms.\\

\noindent
\textbf{Economic Objective:}
Our economic objective to minimize the normalized \emph{excess} outside the bid-ask spread, defined here as
\begin{equation}
  e(C;B,A)
  :=\frac{(C-A)^+ +(B-C)^+}{(A-B)/2}.
  \label{eq:half-spread-excess-preview}
\end{equation}
We report the percentage of options inside the bid-ask spread ($e=0$) and also the mean and conditional excess conditional
on not being within bid-ask.

That means from economic perspective our objective is to minimize the normalized excess outside the bid-ask spread for each quote plus a small
additional regularization term to pull the solution towards the mid. Concretely,
for the $\ell$th quote on date $t$ let $C_t^\ell$ be the model call price, $A^\ell_t$ the ask price, $B^\ell_t$ the bid price, and
$M_t^\ell=(\ A_t^\ell+ B_t^\ell)/2$ the mid price.

Then define our loss as
\begin{equation}
  \mathcal L_t^{\mathrm{quote}}
  :=\frac{1}{L_t}\sum_{\ell=1}^{L_t}
    \frac{
      (C_t^\ell- A_t^\ell)^+
      +(B_t^\ell-C_t^\ell)^+
      +0.05\lvert C_t^\ell-M^\ell_t\rvert
    }{(A_t^\ell - B_t^\ell)/2}.
  \label{eq:production-quote-loss}
\end{equation}

\subsection{Affine Compression}

Considering the large state space we employ an affine compression to reduce dimensionality while retaining most of the relevant information.
This `affine-200' representation is defined by the decoder
\eq{affine-decoder-preview}
u_t=b+B h_t 
\eqend
in terms of learned $h_t\in\R^{200}$, $b\in\R^{M(N+1)+2}$ and
$B\in\R^{(M(N+1)+2)\times200}$.\footnote{
The choice of 200 is an empirical capacity compromise rather than a structural
claim about the intrinsic dimension of option surfaces.  It is large enough
to preserve almost all of the direct native quote fit while remaining below
the number of historical dates, so a full innovation covariance and stable
joint state dynamics can still be estimated.  We report the native-versus-200
fit loss explicitly below rather than treating this dimension as exact.
}
Table~\ref{tab:aggregate-fit-preview} first shows that dimension reduction mostly works: the affine-200 variant costs only 1.36 percentage points
in bid-ask fit.
\begin{table}[H]
\centering
\begin{tabular}{lrrrrrr}
\toprule
Geometry & Inside bid--ask
  & Mean excess & Cond. mean & Cond. median\\
\midrule
Full & 82.28\% & 0.31 & 1.76 & 0.69\\
Affine-200 & 80.92\% & 0.32 & 1.69 & 0.68\\
\bottomrule
\end{tabular}
\caption{Aggregate fit of the affine-200 vs the full model, measured by \emph{excess}~\eqref{eq:half-spread-excess-preview}.}
\label{tab:aggregate-fit-preview}
\end{table}

However, the fit for the one-day to expiry range in the affine-200 model is noticeably worse,
highlighting the challenge of capturing very short-term dynamics.
\begin{table}[H]
\centering
\begin{tabular}{lrrrrr}
\toprule
Geometry & 1D & 2D & 3D & 4D & 5D\\
\midrule
Full & 94.56\% & 95.21\% & 95.57\% & 95.77\% & 68.43\%\\
Affine-200 & {\color{red}88.59\%} & 91.98\% & 93.68\% & 94.13\% & 67.58\%\\
\bottomrule
\end{tabular}
\caption{Inside-spread rates at the five shortest nominal expiries.}
\label{tab:short-fit}
\end{table}

\subsection{Linear Mean-Reverting State-Space Model}
\label{sec:affine-state-space-model}

In the previous section we learned per date ML-SANOS parameters $u_t$
for each $t=1,\dots,t_m$. The next step is to learn a dynamic model. Since we have shown that
we can represent the full $u_t$ sufficiently well using an affine-200 reduction, we will focus on that setup.

Our generative specification learns jointly the affine representation decoder and a stable linear state process:
for $d=200$ let
\begin{align}
  u_t&=b+B h_t,
       &h_t&\in\R^d,
       \label{eq:affine-decoder}\\
  h_{t+\Delta}
     &=m+A_\Delta(h_t-m)+G_\Delta' z_{t,\Delta},
       &A_\Delta&=\exp(-\kappa\Delta),
       \qquad z_{t,\Delta}\sim\mathcal N(0,I_d).
       \label{eq:affine-state-space-transition}
\end{align}
Thus the innovation covariance at horizon $\Delta$ is
$G_\Delta'G_\Delta$.  Here $b,B$ are the affine projections and
$A_\Delta$ is a mean-reversion transition matrix. Mean reversion is imposed
structurally through
\begin{equation}
  S:=\tfrac12(R-R'),\qquad S'=-S,\qquad
  \kappa=LL'+S+\kappa_{\min}I,
  \qquad \kappa_{\min}=0.0002\ \text{per trading session}.
  \label{eq:dissipative-generator}
\end{equation}
Its symmetric part is $LL'+\kappa_{\min}I\succ0$. Therefore every eigenvalue
$\lambda$ of $\kappa$ satisfies $\Re\lambda\geq\kappa_{\min}>0$, and every
eigenvalue of $A_\Delta$ has modulus
$\exp(-\Delta\Re\lambda)<1$.  Hence the transition is stable for every
$\Delta>0$.  This restriction holds during every optimizer update. 

 The transition likelihood is fitted using every pair of observed surfaces separated by exactly
$\Delta=5$ NYSE trading sessions.  We choose this
one-week estimation horizon as a stable
baseline: it attenuates the influence of noisy one-day surface moves while
retaining 861 overlapping transitions for estimation.  Daily simulation uses
the same generator through $A_1=\exp(-\kappa)$ and factors the corresponding
one-session innovation covariance as $G_1'G_1$; it is not a separately fitted
daily AR model.  At the fitted horizon, $G_5$ is a
full upper-triangular matrix whose diagonal is a positive floor plus a
softplus transform.  Consequently $G_5'G_5$ is positive definite throughout
training.

Let $D=891$ and let $\mathcal W$ contain the 861 exact one-week date pairs.
For $(t,s)\in\mathcal W$ with $s=t+\Delta$, set
\eq{oneweekresidual}
  e_{t,s}:=h_s-m-A_\Delta(h_t-m).
\eqend
The Gaussian transition loss per state dimension is
\begin{equation}
  \mathcal L_{\mathrm{AR}}
  :=\frac{1}{2d}\left\{
      \frac{1}{\lvert\mathcal W\rvert}
        \sum_{(t,s)\in\mathcal W}
          \left\|(G_5')^{-1}e_{t,s}\right\|_2^2
      +2\sum_{i=1}^{d}\log (G_5)_{ii}+d\log(2\pi)
    \right\}.
  \label{eq:production-ar-loss}
\end{equation}
With the smooth DLV-cap penalty $\mathcal L_{\mathrm{cap}}$ and affine-gauge
penalty $\mathcal L_{\mathrm{gauge}}$ defined in
Appendix~\ref{app:affine-training}, the jointly trained affine state-space
model uses
\begin{equation}
  \mathcal J_{\mathrm{joint}}
  :=\frac1D\sum_{t=1}^{D}
       \mathcal L_t^{\mathrm{quote}}(b+B h_t)
     +\lambda_{\mathrm{cap}}\mathcal L_{\mathrm{cap}}
     +\lambda_{\mathrm{AR}}\mathcal L_{\mathrm{AR}}
     +\lambda_{\mathrm{gauge}}\mathcal L_{\mathrm{gauge}},
  \label{eq:production-joint-objective}
\end{equation}
where
\[
  \lambda_{\mathrm{cap}}=1,
  \qquad \lambda_{\mathrm{AR}}=0.10,
  \qquad \lambda_{\mathrm{gauge}}=0.10.
\]
The model is deliberately simple.  The affine decoder reduces a
3,620-dimensional arbitrage-free surface coordinate to a state small
enough to support a nonsingular full innovation covariance.  The 
matrix exponential supplies stable transitions at every simulation horizon,
while joint estimation chooses a representation that remains accurate in
quote space and is also statistically coherent through time.  

The implementation minimizes~\eqref{eq:production-joint-objective} over 300
epochs by alternating surface and transition updates, with the AR weight
ramped from zero to its stated value over the first 50 epochs. Table~\ref{tab:aggregate-fit2}
shows that the expanded estimation does not impede overall fit.

\begin{table}[H]
\centering
\begin{tabular}{llrrrrr}
\toprule
Geometry and state & Inside bid--ask
  & Mean excess & Cond. mean & Cond. median\\
\midrule
Full & 82.28\% & 0.31 & 1.76 & 0.69\\
Affine-200 & 80.92\% & 0.32 & 1.69 & 0.68\\
AR(1) affine-200 &\textbf{81.48\%} & 0.32 & 1.72 & 0.68\\
\bottomrule
\end{tabular}
\caption{Aggregate fit before and after the joint AR overlay in terms of \emph{excess}~\eqref{eq:half-spread-excess-preview}.}
\label{tab:aggregate-fit2}
\end{table}

\begin{remark}
  Gaussian innovations miss most of the tail events present in the empirical data. See appendix~\ref{app:tail-dynamics} on page~\pageref{app:tail-dynamics}
  for capturing tail dynamics better.
\end{remark}

\begin{remark}
The 80.92\% and 81.48\% market-fit figures were computed in-sample on the population used to estimate
the final all-history model because the aim here was to build a model which is self-consistent within this population. 

It is not a claim that $b$ and $B$ will maintain the same market-fit out-of-sample.
A future forecasting study should freeze the decoder and dynamics
before the evaluation period.
\end{remark}

\subsubsection{Spot}

The normalized floating option surface can be simulated without considering spot. 
However, for most  applications we will still need a spot path. We will essentially aim to draw
spot from the front expiry marginal distribution. This means that generated spot paths will be consistent with the one day to expiry option prices.
In order to ensure a correct joint distribution between spot and the surface, we use a `copula' approach.

Assume that we have historic samples of daily driftless spot returns $S_{t+1}/S_t$  as defined in~\eqref{eq:driftlessspotret}.  
Let
$F_t^1$ denote the CDF obtained\footnote{For each date we tabulate
$F_t^1$ on a fine log-return grid and invert it with a monotone PCHIP spline.} from the front one-day decoded SANOS density $q_t^1$ and use it 
to transform the spot return into the Gaussian score
\begin{equation}
  dg_t =\Phi^{-1}\!\left[
    F_t^1\!\left(\frac{S_{t+1}}{S_t}\right)
  \right].
  \label{eq:front-spot-score}
\end{equation}

We now simply regress $dg_t$ on the normal $z_{t,1}$ from~\eqref{eq:affine-state-space-transition}. If $\sigma_\perp$ is the fitted residual
standard deviation, we sample a driftless normalized spot path starting in $S_1=1$ using
\begin{equation}
  \bar dg_t=\frac{b'z_{t,1}+\sigma_\perp\epsilon}
          {\sqrt{b'b+\sigma_\perp^2}},
  \qquad
  S_{t+1}
    =S_t\, (F_t^1)^{-1}\!\left(\Phi(\bar dg_t)\right) \ . 
  \label{eq:spot-block}
\end{equation}
Because $F_t^1$ is the option-implied risk-neutral front-expiry marginal, this
construction does not estimate or add a physical spot drift.  

\subsubsection{Time Series Generation}
\label{sec:simulation}

The basic statistical DYSANOS time series generator is summarized as follows:
\begin{enumerate}
  \item Initialize $h_0$ from an observed
        date or from the invariant state distribution.
  \item Propagate the state with the exact Gaussian transition
        \eqref{eq:affine-state-space-transition}, or use the tail diffusion methods described in appendix~\ref{app:tail-dynamics} on
        page~\pageref{app:tail-dynamics}.
  \item Simulate spot with~\eqref{eq:spot-block}.
  \item Form $u_t=b+B h_t$, map it to bounded SANOS parameters through
        \eqref{eq:totalW}--\eqref{eq:defDLVdsq}  and construct the marginal densities
        as described in section~\ref{sec:mlsanos}.
  \item Evaluate $C_t(T,K)$ at any desired strike and expiry. For a fixed 
        cash contract $(\tau,k)$ use $T=\tau-t$ and $K=k/S_t$. Option  prices can  be pre-computed
        if their relative strikes and expiries are known in advance; otherwise 
        they can be evaluated on the path on the fly.\footnote{
          It is much more efficient if the time to expiry of dynamically generated options is known in advance
          as then the respective linear expression is local to the surrounding model expiries. If the
          expiry is entirely free then the price of an arbitrary option in batch-based GPU evaluation is still linear
          but requires without additional tricks evaluating all model options across the entire density surface.
        }
\end{enumerate}

\subsection{SANOS PCA Implied Volatility}
\label{sec:pca}

In this section we review the PCA dynamics implied by our model.

\begin{figure}[H]
  \centering
  \includegraphics[width=0.7\textwidth]{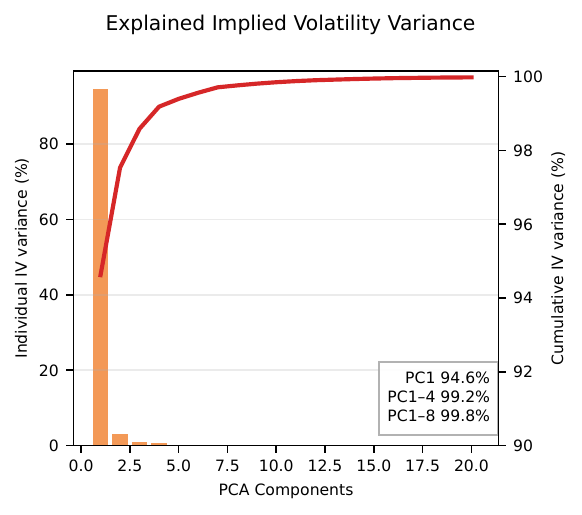}
  \caption{Variance explained by PCA of decoded implied-volatility levels
  under the fitted invariant state law.  The 200 affine coordinates are not
  themselves these PCs.}
  \label{fig:pca-spectrum}
\end{figure}

Although $h_t$ for our affine-200 model has 200 coordinates, the actual variation is
much lower-dimensional.  Indeed, the PCA of the AR(1) state process reveals that only a few principal components capture most of the variation in the decoded
implied-volatility space as figure~\ref{fig:pca-spectrum} and table~\ref{tab:decoded-pca} show. Practically eight PCA vectors are sufficient to drive the entire surface over the
time period of 891 dates from 2022-01-03 to 2025-08-29. This is in line with other works~\citep{ContVol}.

\begin{table}[H]
\centering
\begin{tabular}{lr}
\toprule
Cumulative decoded-IV variance & Explained variance\\
\midrule
PC1 & 94.62\%\\
PC1--2 & 97.53\%\\
PC1--4 & 99.19\%\\
PC1--8 & 99.76\%\\
\bottomrule
\end{tabular}
\caption{Decoded implied-volatility variance under the invariant state law.}
\label{tab:decoded-pca}
\end{table}

\begin{figure}[p]
  \centering
  \includegraphics[width=\textwidth,height=0.88\textheight,keepaspectratio]{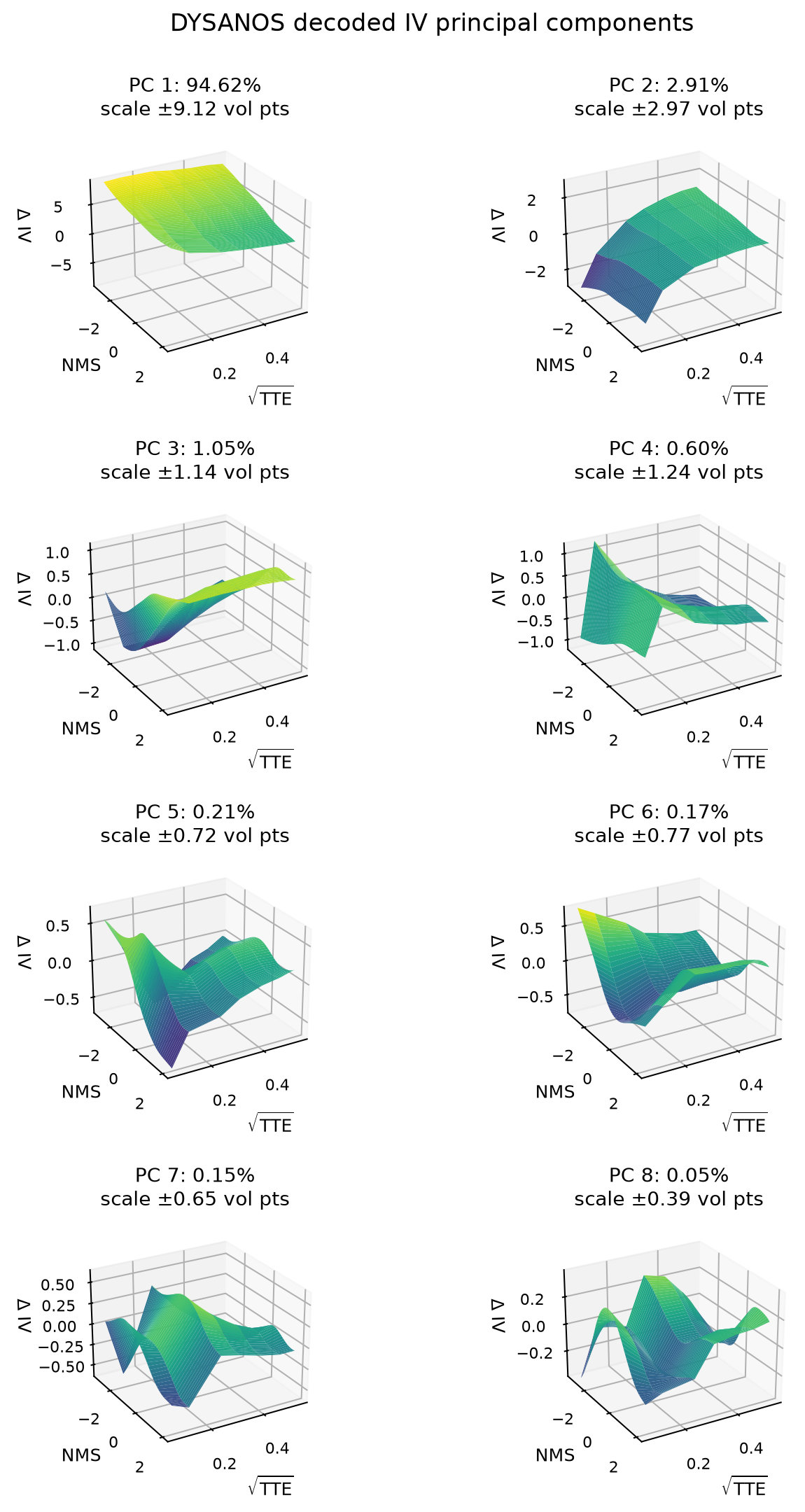}
  \caption{PCA factors 1--8 expressed in implied volatility space as shifts over the mean $b$ of~\eqref{eq:affine-decoder}.}
  \label{fig:pca-factors}
\end{figure}

Figure~\ref{fig:pca-factors} shows the first eight eigensurfaces.  Each
panel is one invariant standard deviation and has its own vertical scale; PCA
signs are arbitrary.  PC1 is the global level.  PC2 is primarily a broad
skew/term-structure rotation, while PC3 couples skew and maturity.  Higher
modes become increasingly localized, including at the explicit 1D--4D model
expiries.

\clearpage

\begin{figure}[H]
  \centering
  \includegraphics[width=0.7\textwidth]{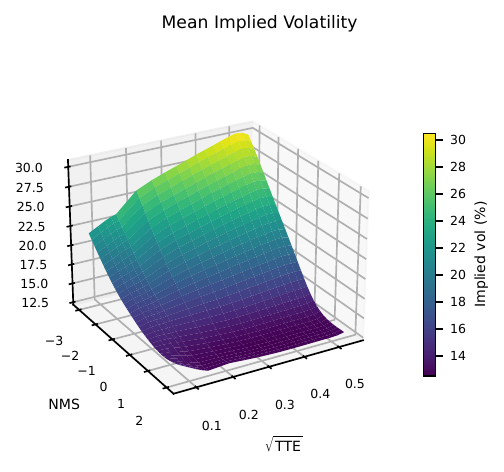}
  \caption{Mean implied-volatility~$b$ from~\eqref{eq:affine-decoder}.}
  \label{fig:base-vol-surfaces}
\end{figure}

For completeness we also illustrate the mean~$b$ from~\eqref{eq:affine-decoder}
in figure~\ref{fig:base-vol-surfaces}.

\subsubsection{Invariant Law and Stability diagnostics}

\begin{table}[H]
\centering
\begin{tabular}{lr}
\toprule
Statistic & Value\\
\midrule
State dimension & 200\\
Dates & 891\\
Exact 1W innovations & 861\\
$\rho(A_5)$ & 0.9561\\
Slowest-mode half-life & 15.4 weeks\\
Minimum one-day real eigenvalue of $\kappa$ & 0.00898\\
Invariant covariance condition number & 3.64\\
Invariant covariance effective rank & 196.4\\
\bottomrule
\end{tabular}
\caption{Fitted state-law diagnostics.}
\label{tab:stability}
\end{table}

The law is stable and retains almost full effective rank in state space.  A
simulator can therefore start from an observed state or draw
$h_0\sim\mathcal N(m,P)$ when a stationary initialization is appropriate.

\subsubsection{Chronological surface-path comparison}
\label{sec:path-realism}

Horizon radii test the latent innovation law but do not directly test paths
of decoded option prices.  We therefore form every available sequence of four
fitted market-state surfaces separated by one full trading week, giving three
consecutive one-week changes.  Here `observed' means the date-specific latent
state fitted to that day's irregular quote book and then passed through the
common SANOS decoder; it is not a second interpolation of the raw quotes.
Every candidate simulator starts from that fitted surface at the same origin
date.  Its path features are changes in normalized call prices on
the Cartesian panel of target locations
\[
  T\in\{1,5,10,20,40,70\}\text{ business days},
  \qquad
  K\in\{0.85,0.90,0.95,1.00,1.05,1.10,1.15\}.
\]
Each target expiry is mapped to the closest distinct model expiry.  The
model therefore has $3\times6\times7=126$ features.  Coordinates are centered
and scaled using pre-2025 observed paths only.

For this path-level test, the historical comparator replays an entire
pre-2025 three-week sequence of fitted residuals, rather than drawing its
three one-week residuals independently.  It is therefore a deliberately strong
finite-sample benchmark that preserves short-run clustering, but it is not a
new parametric law and cannot generate innovations outside the stored blocks.

We report the standardized energy distance and an unbiased Gaussian-kernel
maximum mean discrepancy (MMD) between observed and simulated 2025 paths.
For the null distribution, entire paired chronological blocks of 15 origin
dates swap their observed/simulated labels; the block length matches the
three-week feature horizon and retains most overlap-induced dependence.
1,999 random block swaps give a Monte Carlo $p$-value with minimum positive
resolution $0.0005$.

The 2025 split is a chronological holdout for fitting any auxiliary tail law,
constructing a historical-residual pool, and estimating the feature
standardization.  The shared affine decoder and the baseline Gaussian AR law
are nevertheless taken from the all-history checkpoint.  The path comparison
is therefore a holdout diagnostic for the innovation-law extensions
conditional on the learned representation, not a fully frozen pre-2025 model
forecast.

Table~\ref{tab:path-realism} reports 144 paired 2025 paths.
The $p$-value is the block-swap test for MMD; energy distance is reported as a
second scale-sensitive discrepancy.  None of the five candidates is rejected
at 5\%.  This is evidence against a gross path-distribution failure, but not a
ranking theorem.  Gaussian has the smallest energy distance and MMD.  The
Tail Diffusion remains statistically plausible while improving the
separate tail and persistence diagnostics above.

\begin{table}[H]
\centering
\small
\begin{tabular}{lrrr}
\toprule
Innovation law & Energy & $10^3\!\times$MMD$^2$ & $p$\\
\midrule
Gaussian & 0.810 & 1.594 & 0.704\\
IID Student-$t$ & 0.854 & 3.788 & 0.457\\
Score-driven Student-$t$ & 0.838 & 3.786 & 0.416\\
Tail Diffusion & 0.854 & 3.480 & 0.415\\
Historical block replay & 0.857 & 6.968 & 0.345\\
\bottomrule
\end{tabular}
\caption{Chronological decoded-surface path diagnostics on 144 paired 2025
origins.  Each path contains three consecutive one-week changes; $p$ is based
on 1,999 paired 15-origin block swaps.}
\label{tab:path-realism}
\end{table}

\section{Risk-Neutral DYSANOS}
\label{sec:risk-neutral}

Static arbitrage-freedom of each decoded surface does not by itself impose a
martingale restriction on the option prices. We show that a numerical martingale
measure can nevertheless be constructed for a subset of tradable contracts by
applying the OCE drift-removal construction of
\citet{removedrift} directly to simulated DYSANOS paths.  We do not
claim risk-neutrality of the full surface; extending the construction to
\emph{all} options is ongoing work.

We also stress that the `remove the drift' procedure deployed here is brittle: it is sensitive to numerical
errors, overfitting and other numerical challenges. While we did manage to find a risk-neutral
measure for the paths we generate, it was an iterative process. As it is the method is not suitable
for a production environment.

\subsection{Universe}
\label{sec:risk-neutral-universe}

We simulate 25 trading dates over
$N=100{,}000$ paths under the learned measure~$\P$.
The paths are generated antithetically: at every daily
transition we draw the state and independent spot innovations for the first
half of the block and use their exact negatives for the second half.  The two
members are then propagated through the same nonlinear decoder.
Consequently, the sample contains $N/2=50{,}000$
independent antithetic pairs rather than $N$ independent paths.\footnote{The antithetic pairing matters when estimating
their Monte Carlo uncertainty.}

At each date we can trade spot and 12 floating options
at each of the
four expiries $5D,10D,15D,20D$ struck on the  normalized-moneyness
grid
\[
z\in\{-1.5,-1.3,-1.1,-0.9,-0.7,-0.5,-0.3,-0.1,0,0.1,0.3,0.5\}.
\]
Once entered, each option is held to expiry.
Writing $\Pi_t^i$ for the (spot or option) entry price and $X_t^i$ for the (spot or option) expiry value of instrument
$i$, returns are expressed as fractional gains:
\begin{equation}
  R_t^i=\frac{X_t^i}{\Pi_t^i}-1,
  \qquad i=1,\ldots,49 \ .
  \label{eq:rn-returns}
\end{equation}
This normalizes our trading to unit \$ amounts invested in each tradable instrument. We only trade ITM options to avoid vanishing premia.

\begin{remark}
We are going to learn a measure for the \emph{fixed} 100,000 paths we have generated. It is not a general function of the state that would
generalize to other generated paths. We are looking for an in-sample risk-neutral measure.
Therefore, we do not split training and validation set for the below analysis.

A different task is to find a generic risk-neutral measure for a trained DYSANOS model. This is not the focus of the current analysis.
\end{remark}

\subsection{Finding a Risk Neutral Measure}

As a first step we find a pre-conditioning measure $\Q_1$ equivalent to $\P$ under which the unconditional expectation of every return is zero to numerical tolerance.
Let $D$ be the density of $\Q_1$ with respect to~$\P$ over $\R^N$. Then we solve\footnote{
  Numerically, we solve the smooth unconstrained dual
  $\inf_a\log[N^{-1}\sum_n\exp(-R_n'a)]$ using trust-region Newton--CG with analytic gradients and exact covariance Hessian--vector products.
  Float64 log-sum-exp stabilization is used for the objective and gradient, after which $D_n=\exp(-R_n'a)/[N^{-1}\sum_m\exp(-R_m'a)]$ is reconstructed;
  hence every weight is strictly positive and their sample mean is one.}
\eq{convex}
  \inf_{D\in \R^N} \E^\P\left[ D\, \log D \right]
\eqend
subject to the linear constraints
\eq{convexcond}
  D > 0 \ , \ \ \ \E^\P\left[ D \right ] = 1 \ \ \ \mbox{and} \ \ \ \E^\P\left[ D\,R_t \right] = 0
\eqend
Table~\ref{tab:rn-precondition} shows summary results of this step.
\begin{table}[H]
  \centering
  \begin{tabular}{@{}ll@{}}
    \toprule
    Diagnostic & Reported value \\
    \midrule
    Simulated paths & $ 100{,}000$ \\
    Returns & $1{,}225=25\times49$ \\
    Moment constraints & exact zero \\
    Sample relative entropy & $0.2011$ \\
    Effective sample size & $62{,}877$ ($62.88\%$ of paths) \\
    Minimum / maximum density & $0.0019$ / $44.53$ \\
    Maximum unconditional $|z|$ & $6.85\times10^{-9}$ \\
    \bottomrule
  \end{tabular}
  \caption{Entropy pre-conditioning,
  evaluated on the 100,000-path sample used in the diagnostics below.}
  \label{tab:rn-precondition}
\end{table}

The next step is to construct a measure which has numerically zero conditional
expected returns, i.e., a risk-neutral measure.  To this end, recall the
Minimal Equivalent Martingale Measure (MEMM)~\citep{MEMM}, which is a special
case of the construction in~\citet{removedrift}:

\begin{theorem}
Assume our option/spot returns $R$ are $L^1(\Q_1)$ and let $\calA$ be the set of bounded variables $a=(a_0,\ldots,a_{24})$
each with values in  $\R^{49}$, measurable with respect to the filtration generated by the market.
Solve
\eq{rn-oce-general}
  \sup_{a\in\calA}:
  -\log\E^1\left[
    \exp\left(
      -\sum_{t=0}^{24}\left(a_t'R_t-|a_t|'c_t\right)
    \right)
  \right]
  \longrightarrow a^0
\eqend
If $a^0$ is finite, then
\eq{densit1}
  \frac{d\Q_2}{d\Q_1}
  :=\frac{
    \exp\left[-\sum_{t=0}^{24}\left({a_t^0}'R_t-|a_t^0|'c_t\right)\right]
  }{
    \E^1\left[
      \exp\left[-\sum_{t=0}^{24}\left({a_t^0}'R_t-|a_t^0|'c_t\right)\right]
    \right]
  }
\eqend
defines an equivalent measure.
\end{theorem}

In our example, we parameterize the action set $a$ as follows:
for expiry $k$, let
\[
  x_{t,k}=(t,h_t,\log\Pi_t^{k,1},\ldots,\log\Pi_t^{k,12}).
\]
The action network has a two-hidden-layer width-128 SiLU component
$F_\theta:\R^{213}\to\R^{13}$ shared across expiries and an
expiry-specific affine skip $b_{\theta,k}+W_{\theta,k}x_{t,k}$.  Writing
\begin{equation}
  r_{\theta,t,k}=b_{\theta,k}+W_{\theta,k}x_{t,k}
    +F_\theta(x_{t,k})\in\R^{13},
  \qquad k=1,\ldots,4,
  \label{eq:rn-expiry-action}
\end{equation}
the four first coordinates are summed into one spot action and the remaining
coordinates give the 48 option actions.  This defines
$a_\theta(x_t)\in\R^{49}$ with $\Theta=56{,}709$ parameters.  Thus the
strategy uses the full 200-dimensional state, all option entry prices, and
calendar time.

Let $c_t$ contain one basis point for spot and ten basis points for every
option.  The $\Q_2$ fit uses $c_t^{(2)}=c_t/10$, i.e., $0.1$ basis points for
spot and one basis point for every option, together with the capped action
\begin{equation}
  \bar a_\theta(x_t)
  =a_\theta(x_t)
   \min\!\left(1,\frac{32}{\lVert a_\theta(x_t)\rVert_1}\right) \ .
  \label{eq:rn-action-cap}
\end{equation}
We then solve the problem
\begin{equation}
  \sup_{\theta\in\R^\Theta}: \
  -\log\E^1\!\left[
    \exp\!\left(
      -\sum_{t=0}^{24}\left\{
        \bar a_\theta(x_t)'R_t
        -\lvert\bar a_\theta(x_t)\rvert'c_t^{(2)}
      \right\}
    \right)
  \right]
  \;\longrightarrow\;\theta^0.
  \label{eq:rn-oce-q2-fit}
\end{equation}
on the full 100,000 paths, using an exact full-population Adam warm start
followed by exact full-population L-BFGS.  The full objective and exact-gradient
history is reported in Appendix~\ref{app:rn-q2-convergence}.
Our candidate martingale measure is therefore
\begin{equation}
  \frac{d\Q_2}{d\Q_1}
  =\frac{
    \exp\!\left(
      -\sum_t\left\{
        \bar a_{\theta^0}(x_t)'R_t
        -\lvert\bar a_{\theta^0}(x_t)\rvert'c_t^{(2)}
      \right\}
    \right)
  }{
    \E^1\!\left[
      \exp\!\left(
        -\sum_t\left\{
          \bar a_{\theta^0}(x_t)'R_t
          -\lvert\bar a_{\theta^0}(x_t)\rvert'c_t^{(2)}
        \right\}
      \right)
    \right]
  } \ .
  \label{eq:rn-q2-density}
\end{equation}

\subsection{No-Arb Test}

We deploy two types of tests for the no-arbitrage condition: unconditional and conditional.

\subsubsection{Unconditional Test}

Every unconditional day--instrument mean is compared with its Monte Carlo
standard error.  We reconstruct the $\Q_1$ and terminal cap-32 $\Q_2$
densities on the same 100,000 paths, split into sets~$A$ and~$B$ while keeping
antithetic pairs together. 
For each $\Q\in\{\P,\Q_1,\Q_2\}$, write the self-normalised weights as
$w_p=d\Q/d\P(p)$.  Their effective sample sizes are 100,000, 71,175, and
33,380, respectively.  For each of the $25\times49=1{,}225$ coordinates define
\begin{equation}
  \widehat m_{t,i}
  =\frac{\sum_p w_pR_{p,t}^i}{\sum_p w_p},
  \qquad
  z_{t,i}
  =\frac{\widehat m_{t,i}}
  {\widehat{\operatorname{se}}(\widehat m_{t,i})}.
  \label{eq:rn-unconditional-z}
\end{equation}
The standard error is the self-normalised ratio-estimator influence standard
error after averaging the two paths in each antithetic pair.  We also report
the materiality-regularised score
\begin{equation}
  s_{t,i}^{(\varepsilon)}
  =\frac{\widehat m_{t,i}}
  {\sqrt{\widehat{\operatorname{se}}(\widehat m_{t,i})^2+\varepsilon^2}},
  \qquad \varepsilon=10^{-8}.
  \label{eq:rn-materiality-score}
\end{equation}

\begin{table}[H]
  \centering
  \small
  \begin{tabular}{@{}lrrr@{}}
    \toprule
    Diagnostic & $\P$  & $\Q_1$  & $\Q_2$  \\
    \midrule
    Paths / antithetic pairs
      & $100{,}000$ / $50{,}000$ & $100{,}000$ / $50{,}000$
      & $100{,}000$ / $50{,}000$ \\
    Effective sample size & $100{,}000$ & $71{,}175$ & $33{,}380$ \\
    Day--instrument coordinates & $1{,}225$ & $1{,}225$ & $1{,}225$ \\
    Maximum $|z|$ & $53.026$ & $2.205$ & $2.646$ \\
    99th percentile $|z|$ & $45.706$ & $2.032$ & $1.833$ \\
    $N_{|z|>1.96}$ & $1{,}131$ & $19$ & $9$ \\
    $N_{|z|>2.576}$ & $1{,}096$ & $0$ & $1$ \\
    $N_{|z|>5}$ & $984$ & $0$ & $0$ \\
    Maximum $|s^{(\varepsilon)}|$ & $53.026$ & $2.205$ & $2.646$ \\
    $N_{|s^{(\varepsilon)}|>1.96}$ & $1{,}131$ & $19$ & $9$ \\
    $N_{|s^{(\varepsilon)}|>2.576}$ & $1{,}096$ & $0$ & $1$ \\
    $N_{|s^{(\varepsilon)}|>5}$ & $984$ & $0$ & $0$ \\
    \bottomrule
  \end{tabular}
  \caption{Unconditional coordinate diagnostics under $\P$, the
  pre-conditioned measure $\Q_1$, and the risk-neutral measure $\Q_2$.
  All three columns use the same paths; antithetic pairs are the independent
  sampling units.}
  \label{tab:rn-q2-unconditional}
\end{table}

Under~$\P$, 984 coordinates exceed five standard errors; the worst is the
day-1 20D $z=-0.7$ ITM call return, with mean $-0.0540420$, standard error
$0.0010192$, and $z=-53.026$.  Under~$\Q_1$, no coordinate exceeds 2.576
standard errors; its worst coordinate is the day-1 spot return, with mean
$0.0000636$, standard error $0.0000288$, and $z=2.205$.  Under~$\Q_2$, one
coordinate exceeds 2.576 standard errors and
none exceeds five; its worst coordinate is the day-4 spot return, with mean
$-0.0001772$, standard error $0.0000670$, and $z=-2.646$.  Coordinate
dependence means that the counts are diagnostics rather than family-wise
$p$-values.

\subsubsection{Conditional Test}

As a next step, we test whether the conditional returns of our trading instruments have zero mean.
This is complicated by the nature of non-recombining Monte-Carlo simulated paths where no state (practically) appears twice.

We therefore propose a novel conditional no-arbitrage test:

\begin{theorem}[Conditional No-Arb Test]\label{th:cond-no-arb}
Let $\Q$ be a measure on a filtered space
$(\Omega,\mathcal{F}_0,\ldots,\mathcal{F}_m)$ and let
$R_t\in L^1(\Q;\R^d)$, $t=0,\ldots,m-1$.  Let $\mathcal A$ contain all bounded
$\mathcal F_t$-measurable actions $a_t$.  Assume the conditional Laplace
transforms used below are finite and twice differentiable on a convex
neighbourhood containing zero and the optimizer.  If
$a^*=(a_0^*,\ldots,a_{m-1}^*)$ is a finite solution of
\begin{equation}
  \inf_{a\in\mathcal A}: \E^\Q\!\left[
    \sum_{t=0}^{m-1}
    \exp\!\left(-a_t'R_t+\frac12a_t'a_t\right)
  \right],
  \label{eq:population-action-objective}
\end{equation}
then, for every $t$,
\begin{equation}
  \E^\Q[{a_t^*}'R_t]=0
  \quad\Longleftrightarrow\quad
  \E_t^\Q[R_t]=0\quad\Q\text{-a.s.}
  \label{eq:population-action-equivalence}
\end{equation}
\end{theorem}
\noindent
(For a proof refer to the appendix page~\pageref{sec:proof-cond-no-arb}.)

Accordingly, we first split our 100,000 paths into two sets $A$ and $B$, each containing
50,000 paths (25,000 antithetic pairs).  For each $\Q\in\{\P,\Q_1,\Q_2\}$ fit
an action with the full affine-plus-deep architecture
in~\eqref{eq:rn-expiry-action}, but with no action cap, by minimizing\footnote{
Features are standardized using
the weighted training fold only; both folds and measures use the same
300-epoch schedule: 100 affine-initialization epochs followed by 200
full-network epochs.}
\begin{equation}
  \widehat L_{A,\Q}(\phi)
  =\log\!\left\{
    \frac1{25}\widehat\E_A^{\Q}\!\left[
      \sum_{t=0}^{24}
      \exp\!\left(
        -a_\phi(x_t)'R_t+\frac12a_\phi(x_t)'a_\phi(x_t)
      \right)
    \right]
  \right\}.
  \label{eq:rn-conditional-objective}
\end{equation}  For conditionally
Gaussian returns, the population action is approximately
$(I+\operatorname{Cov}_t(R))^{-1}\E_t^{\Q_j}[R]$.

After fitting on $A$, the action is evaluated on $B$ using
\begin{equation}
  d_{p,A\to B}^{(j)}
  =\frac1{25}\sum_{t=0}^{24}
    a_{\widehat\phi_{A,j}}(x_{p,t})'R_{p,t}.
  \label{eq:rn-holdout-score}
\end{equation}
We then exchange $A$ and $B$ and pool the two out-of-fold score vectors.  The
combined mean is
\begin{equation}
  \widehat m_{\mathrm{OOF}}^{(j)}
  =\frac{
    \sum_{p\in B}w_p d_{p,A\to B}^{(j)}
    +\sum_{p\in A}w_p d_{p,B\to A}^{(j)}
  }{
    \sum_{p\in A\cup B}w_p
  }.
  \label{eq:rn-oof-mean}
\end{equation}
Let $d_p^{(j)}$ denote the pooled held-out score and
$\bar w=N^{-1}\sum_p w_p$.  For an antithetic pair $k$, define its
self-normalised ratio influence by
\begin{equation}
  \bar\psi_k^{(j)}
  =\frac12\sum_{p\in k}
    \frac{w_p}{\bar w}
    \left(d_p^{(j)}-\widehat m_{\mathrm{OOF}}^{(j)}\right)\ ,
  \qquad
  \widehat{\operatorname{se}}_{\mathrm{pair}}^{(j)}
  =\frac{\operatorname{sd}_{k=1,\ldots,50{,}000}
    (\bar\psi_k^{(j)})}{\sqrt{50{,}000}}\ .
  \label{eq:rn-oof-se}
\end{equation}
The reported action statistic and two-sided normal $p$-value are
\begin{equation}
  T_{\mathrm{act}}^{(j)}
  =\frac{\widehat m_{\mathrm{OOF}}^{(j)}}
    {\widehat{\operatorname{se}}_{\mathrm{pair}}^{(j)}} \ ,
  \qquad
  p_{\mathrm{act}}^{(j)}
  =2\Phi\!\left(-\left|T_{\mathrm{act}}^{(j)}\right|\right)
  \label{eq:rn-action-test}
\end{equation}
where $\Phi$ is the standard-normal distribution function.  Under the null
of zero conditional drift, $T_{\mathrm{act}}^{(j)}$ is asymptotically
standard normal.

\begin{table}[H]
  \centering
  \small
  \begin{tabular}{@{}lrrr@{}}
    \toprule
    Diagnostic & $\P$  & $\Q_1$  & $\Q_2$  \\
    \midrule
    Paths / antithetic pairs
      & $100{,}000$ / $50{,}000$ & $100{,}000$ / $50{,}000$
      & $100{,}000$ / $50{,}000$ \\
    $T_{\mathrm{act}}$ & $93.414$ & $41.365$ & $-0.454$ \\
    Two-sided $p$-value
      & $10^{-1897}$ & $10^{-374}$ & $0.650$ \\
    \bottomrule
  \end{tabular}
  \caption{Conditional zero test, confirming that $\Q_2$ is numerically a martingale measure.}
  \label{tab:rn-q2-conditional}
\end{table}

\removeblock{
\begin{table}[H]
\centering
\small
\begin{tabular}{lrrrrrr}
\toprule
Measure & Max. $|z|$ & 99\% $|z|$ & $N_5$
  & Max. $|s^{(\varepsilon)}|$
  & 99\% $|s^{(\varepsilon)}|$ & $T_{\mathrm{act}}$\\
\cmidrule(lr){2-6}
 & \multicolumn{5}{c}{Unconditional} & \\
\midrule
$\Q_1$ & 2.205 & 2.032 & 0 & 2.205 & 2.032
  & {\color{red}41.365}\\
$\Q_2$ & {\color{orange}2.646} & 1.833 & 0
  & {\color{orange}2.646} & 1.833 & $-0.454$\\
\bottomrule
\end{tabular}
\caption{Compact 100,000-path DYSANOS diagnostics.  The five unconditional
columns summarize the 1,225 day--instrument coordinates, with
$N_5=\#\{|z|>5\}$ and $\varepsilon=10^{-8}$; the last column is the two-fold
out-of-fold conditional action statistic.  The materiality floor is negligible
at the reported coordinates, so $z$ and $s^{(\varepsilon)}$ agree to three
decimal places.}
\label{tab:rn-q-summary}
\end{table}
}

\begin{table}[H]
\centering
\small
\begin{tabular}{lrrrrrr}
\toprule
Measure & $N$ & Max. $|z|$ & 99\% $|z|$
  & $N_{2.576}$ & $N_5$ & $T_{\mathrm{act}}$\\
\cmidrule(lr){3-6}
 & & \multicolumn{4}{c}{Unconditional} & \\
\midrule
$\Q_1$ & $1{,}225$ & 2.205 & 2.032 & 0 & 0
  & {\color{red}41.365}\\
$\Q_2$ & $1{,}225$ & {\color{orange}2.646} & 1.833 & 1 & 0
  & $-0.454$\\
\bottomrule
\end{tabular}
\caption{Threshold-count summary of the 100,000-path DYSANOS diagnostics.
Here $N$ is the number of day--instrument coordinates and
$N_c=\#\{|z|>c\}$. Further confirmation that~$\Q_2$ is a martingale measure.}
\label{tab:rn-q-threshold-summary}
\end{table}

The conditional mean test fails clearly for $\P$ and $\Q_1$, but passes comfortably for $\Q_2$.

\begin{remark}
	Solving~\eqref{eq:rn-oce-q2-fit} is \emph{slow}. In our examples took several hours on an RTX~5090 to find a measure~$\Q_2$.
	That means with the given L-BFGS solver this method is not practically useful if a result is required in minutes or less.
\end{remark}

\subsection{Scope and limitations}
\label{sec:risk-neutral-scope}

This is a reduced demonstration rather than a general construction.  The
experiment is numerically heavy and the terminal $\Q_2$ policy and conditional
actions did not meet strict full-parameter gradient stopping criteria within
the fixed 300-epoch two-stage schedule.

The result is therefore evidence for zero drift within the stated
49-instrument book and strategy class, but it is by construction not a risk-neutrality statement for
other strikes, expiries, trading frequencies, or the full DYSANOS surface.
The model-free reweighting step also relies on having many more simulated paths
than tradable instruments.  Extending the construction to the full surface is
the natural next step.

As part of our implementation work we had to manually intervene to adjust
the optimization process to obtain the desired density. This means the method described
here can serve as a research tool, but is not a practical method for production use. A more robust
approach is subject to further research.

\section{Conclusion}

We presented with DYSANOS the first generative option market model built from
smooth, strictly statically arbitrage-free SANOS surfaces. Our approach does not rely on
approximate static constraints; each smooth option surface is strictly arbitrage free and allows
pricing of any option on GPU.

We have shown how
to leverage our representation to build a simple, but already fairly powerful baseline
model which can generate spot and option paths for many days in the future up to years.
The surface-first AR(1) captures first-order level, skew, term, and leverage
effects.  Its front and term smoothing coordinates allow the short end to move
separately from later expiries, and the affine model places
80.92\% of the full quote population inside bid--ask across all 891 historic dates. The Tail Diffusion
improves the deficient Gaussian
tail law and reproduces tail persistence, although multi-week tail mass
remains too low.  

For a reduced 49-instrument book on the same simulated paths, the MEMM
construction in Section~\ref{sec:risk-neutral} builds a candidate
martingale measure~$\Q_2$.  In the 100,000-path unconditional test the
largest coordinate statistic is $|z|=2.646$, with no coordinate above five.
The two-fold uncapped affine-plus-deep test gives $T_{\mathrm{act}}=41.365$
under $\Q_1$ and $T_{\mathrm{act}}=-0.454$ ($p=0.650$) under $\Q_2$.  We
therefore detect no residual conditional drift under $\Q_2$ in the tested book
and strategy class.  This remains a reduced, computationally heavy
demonstration rather than a full-surface risk-neutral construction.

\section*{Acknowledgements}

We want to thank the team at Wharton for their WRDS \url{https://wrds-www.wharton.upenn.edu} portal and Option Metrics \url{https://optionmetrics.com/united-states} 
for providing their IvyDB data via this service. Without this data our work would not have been possible.

\section*{Declaration of Funding}

B.~Horvath thankfully acknowledges financial support from UKRI1010: High order mathematical and
computational infrastructure for streamed data that enhance contemporary generative and large language models.\\

\noindent
A.~Kratsios gratefully acknowledges financial support from an NSERC Discovery Grant No. RGPIN-2023-04482 and No. DGECR-2023-00230.

\section*{Disclosure of Interest}

There are no interests to declare.

\section*{Declaration of generative AI use}

OpenAI Codex (GPT-5) was used to assist with software implementation,
testing and numerical verification of the models reported in this article,
and with copy-editing and language refinement of author-written text. The
research design, mathematical arguments, interpretation of results and final
manuscript were determined and approved by the authors. The authors verified
the originality and accuracy of all generated code, numerical results,
references and text, checked the tool's terms of use and confirmed its
suitability for publication, and take full responsibility for the integrity
of the article.

\section*{Data availability statement}

The market data supporting the findings of this study are proprietary
OptionMetrics IvyDB data accessed through WRDS. They are available from
OptionMetrics under licence, and restrictions apply to their redistribution.

\bibliographystyle{plainnat}
\bibliography{references}

\clearpage
\appendix

\section{Training the affine state-space model}
\label{app:affine-training}

This appendix provides details on the estimation sequence behind the three rows of model
comparisons in Table~\ref{tab:aggregate-fit2}.  All stages use the same 891
market dates and equal total weight per date, irrespective of the number of
quotes observed on that date.

\subsection{Direct surface benchmark}

For each date, the direct benchmark optimizes an independent
$u_t\in\R^{M(N+1)+2}$ against~\eqref{eq:production-quote-loss}.  The raw
log-DLV coordinates are unbounded above, whereas pricing mechanically applies
the cap in~\eqref{eq:defDLVdsq}.  To retain a restoring gradient around and
above that cap, define
\begin{equation}
  u_{\max}^{\Sigma}:=\log(\Sigma_{\max}-\Sigma_{\min}),
  \qquad a:=0.10,
  \qquad s:=0.35,
\end{equation}
and
\begin{equation}
  r_t^{j,i}
  :=\frac{a}{s}\log\!\left[
       1+\exp\!\left(
         \frac{u_t^{\Sigma,j,i}-u_{\max}^{\Sigma}}{a}
       \right)
     \right].
\end{equation}
The full-data cap penalty is
\begin{equation}
  \mathcal L_{\mathrm{cap}}
  :=\frac1D\sum_{t=1}^{D}
       \frac12\sum_{j=1}^{M}\sum_{i=1}^{N}(r_t^{j,i})^2.
  \label{eq:dlv-cap-penalty}
\end{equation}
Far above the cap this approaches a quadratic penalty in the excess raw
log-DLV coordinate with scale $s$; $a$ determines the smooth transition.  The
direct benchmark minimizes the equal-date quote loss plus
$\mathcal L_{\mathrm{cap}}$.

\subsection{Static affine-200 dimension reduction}

The direct states $u_t^\star$ provide only an initialization for the affine
model.  An analytic rank-200 singular-value decomposition first solves the
least-squares reconstruction problem
\[
  \min_{b,B,\{h_t\}}
  \sum_{t=1}^{D}\lVert u_t^\star-(b+B h_t)\rVert_2^2.
\]
The resulting $b$, $B$, and $h_t$ are then fine-tuned directly against the
market objective
\begin{equation}
  \frac1D\sum_{t=1}^{D}
    \mathcal L_t^{\mathrm{quote}}(b+B h_t)
  +\mathcal L_{\mathrm{cap}}.
  \label{eq:static-affine-objective}
\end{equation}
The reconstruction loss to $u_t^\star$ is not retained during this phase.
The static affine-200 results in Table~\ref{tab:aggregate-fit2} are therefore
market-objective fits, not untouched PCA reconstructions.  This phase contains
no AR likelihood.

\subsection{Exact affine re-gauging and AR initialization}

Before fitting dynamics, the completed affine coordinates are centered and
whitened by an exact change of coordinates.  The decoder bias and loading are
transformed at the same time so that every decoded state $u_t=b+B h_t$ is
unchanged to numerical precision.  Hence this operation changes neither a
surface nor a fitted option price.

For the re-gauged states, every origin date whose endpoint lies exactly five
NYSE sessions later supplies one pair $(h_t,h_s)$.  There are 861 such
overlapping pairs.  Ordinary least squares supplies an initial five-session
transition $A_5$.  Its first-order generator proxy is rescaled to per-session
units; the symmetric part is projected onto the dissipative cone while its
skew-symmetric part is retained.  The initial innovation covariance is the
population covariance of the resulting one-week residuals, with a
$10^{-6}I$ numerical ridge.  Its upper Cholesky factor initializes $G_5$.
This produces a stable warm start for
\eqref{eq:affine-state-space-transition}; it is not yet the final joint fit.

\subsection{Joint AR overlay}

The joint stage runs for 300 epochs and alternates two kinds of optimizer
steps.
\begin{enumerate}
  \item The \emph{surface steps} traverse all dates in shuffled batches and
        update $b$, $B$, and the corresponding $h_t$ using the quote loss and
        DLV-cap penalty.
  \item One \emph{transition step} uses all 861 exact-five-session pairs and updates
        $h_t$, $m$, the dissipative generator parameters, and the full
        upper-triangular innovation factor $G_5$ using
        $\mathcal L_{\mathrm{AR}}$ and the gauge penalty below.  The decoder
        $b,B$ receives no direct gradient from this transition step.
\end{enumerate}
The dynamics weight is ramped linearly over the first 50 epochs.  The state,
decoder, and dynamics learning rates start at $10^{-4}$, $3\times10^{-5}$,
and $3\times10^{-4}$ respectively, after a 20-epoch warm-up, and follow a
cosine schedule to one tenth of those values.  Gradient norms are capped at
10.

For $\bar h=D^{-1}\sum_t h_t$, define the population covariance
\begin{equation}
  \widehat\Gamma_h
  :=\frac1D\sum_{t=1}^{D}(h_t-\bar h)(h_t-\bar h)'.
\end{equation}
The affine-gauge penalty is
\begin{equation}
  \mathcal L_{\mathrm{gauge}}
  :=\frac1d\lVert\bar h\rVert_2^2
    +\frac1d\lVert\widehat\Gamma_h-I_d\rVert_F^2.
  \label{eq:affine-gauge-penalty}
\end{equation}
It controls translations, scale, and shear of the otherwise non-identifiable
affine coordinates; orthogonal rotations and reflections remain immaterial.
The joint stage uses the weights stated below
\eqref{eq:production-joint-objective}.  All 891 dates enter both its surface
and transition objectives, so its fit and dynamics diagnostics are in-sample.

The unrestricted direct state was not given an analogous full-covariance
joint AR overlay.  Its dimension is 3,620, exceeding the 861 observed
one-week transitions.  A nonsingular unrestricted innovation covariance is
therefore not identified without another dimension-reduction or covariance
model.

\section{Tail dynamics}\label{app:tail-dynamics}

This section provides alternatives to the Gaussian diffusion of $h_{t+\Delta}$
in~\eqref{eq:affine-state-space-transition} aimed to better capture tail behavior.

Gaussian innovations are a convenient baseline but miss market tails.  As a
non-parametric comparator we also simulate at one-week steps by independently
resampling centered fitted one-week residuals $e_{t,s}$ from~\eqref{eq:oneweekresidual} whose endpoints precede 2025.
At two and three weeks, independently drawn residuals are propagated through
$A$ and summed; no multi-week block is replayed in this tail experiment.

For an exact-horizon state return $r$ with innovation factor $G_H$,
define its normalized squared radius
\begin{equation}
  (\delta^H)^2:=\frac{\left\|(G_H')^{-1}r\right\|_2^2}{200}.
  \label{eq:radius}
\end{equation}
The Gaussian reference would concentrate this quantity sharply around one in
200 dimensions.  The observed radii are both heavier-tailed and serially
dependent.  Table~\ref{tab:gaussian-historical-tail-gap} makes the scale of
the discrepancy explicit.

\begin{table}[H]
\centering
\small
\begin{tabular}{lrrr}
\toprule
Innovation law & 1W & 2W & 3W\\
\midrule
Gaussian & 0.03\% & $<0.01\%$ & $<0.01\%$\\
Independent historical replay & 5.15\% & 2.41\% & 2.32\%\\
Historical/Gaussian ratio & $162.01\times$ & $17{,}197.36\times$
  & $17{,}430.29\times$\\
\bottomrule
\end{tabular}
\caption{Probability of exceeding the pre-2025 empirical 95\%
normalized-radius threshold under Gaussian innovations and independent
historical resampling.  Ratios use the unrounded probabilities; each simulated
estimate uses one million paths.}
\label{tab:gaussian-historical-tail-gap}
\end{table}

To obtain a more realistic tail law, we add a scalar latent log-variance
diffusion without replaying any historical residual.  Keeping the affine conditional mean and loading $G$
from the fitted factor $G_5$, the exact one-week transition is
\begin{align}
  \varepsilon_{n+1}
    &=\sqrt{\rho\exp(g_n-v_g/2)}\,G_5'z_{n+1},
      &z_{n+1}&\sim\mathcal N(0,I),\label{eq:sv-innovation}\\
  g_{n+1}
    &=\phi g_n+\sqrt{v_g(1-\phi^2)}\,\xi_{n+1},
      &\xi_{n+1}&\sim\mathcal N(0,1),\label{eq:sv-logvariance}
\end{align}
with independent $z_{n+1}$ and $\xi_{n+1}$.  Stationarity gives
$g_n\sim\mathcal N(0,v_g)$ and hence
$\E[\rho\exp(g_n-v_g/2)]=\rho$.  Equation
\eqref{eq:sv-logvariance} is the exact discrete transition of a stationary
Ornstein--Uhlenbeck diffusion; it therefore permits fractional-time simulation
without Euler discretization.

The parameters $(\rho,\phi,v_g)$ are fitted on the 703 pre-2025 one-week
innovations.  Since
\begin{equation}
  \log (\delta_n^2)
  =\log\rho+g_n-v_g/2+\log(\chi^2_{200}/200),
  \label{eq:sv-log-radius}
\end{equation}
we use a Kalman likelihood whose measurement mean and variance are the exact
first two moments of $\log(\chi^2_{200}/200)$; only its Gaussian shape is an
approximation.  The fit gives
\[
  (\rho,\phi,v_g)=(0.878,0.139,0.0649).
\]
On the 158 2025 innovations,
log-radius negative log likelihood falls from 2.453 to 0.125 relative to a
pre-2025-scaled Gaussian benchmark.

For comparison we also fit variance-normalized multivariate Student-$t$
innovations, either IID or with a persistent score-driven common scale.  The
fitted degrees of freedom are 34.5.  On the same 2025 innovations the
score-driven law reduces full-vector negative log likelihood per dimension
from 1.351 to 1.332; the improvement over IID Student-$t$ is negligible.

Table~\ref{tab:tails} reports the probability of exceeding the pre-2025
empirical 95\% radius at each horizon.  `Market' is the realized 2025 rate;
every simulated estimate uses one million paths.

\begin{table}[H]
\centering
\small
\begin{tabular}{lrrr}
\toprule
Innovation law & 1W & 2W & 3W\\
\midrule
2025 market & 6.33\% & 6.33\% & 5.70\%\\
Gaussian & 0.03\% & $<0.01\%$ & $<0.01\%$\\
IID Student-$t$ & 4.14\% & 0.99\% & 0.86\%\\
Score-driven Student-$t$ & 4.12\% & 1.07\% & 0.93\%\\
Tail Diffusion & 3.75\% & 0.78\% & 0.66\%\\
Independent historical replay & 5.15\% & 2.41\% & 2.32\%\\
\bottomrule
\end{tabular}
\caption{Probability of exceeding the pre-2025 market 95\% normalized-radius
threshold.  Market frequencies use 158 2025 transitions; simulation uses one
million paths per law.}
\label{tab:tails}
\end{table}

The Tail Diffusion is an improvement over the Gaussian law. 
 It produces a squared-radius serial correlation of 0.116,
compared with 0.091 in the market history; Gaussian and IID laws have zero
population correlation. It has, however, somewhat less tail mass than IID Student-$t$ at every horizon.\footnote{The 2025
sample is small and overlapping: 15-origin circular-block bootstrap 95\%
intervals for the market exceedance rates are approximately 0.6--14.6\%
across the three horizons.  Differences of a percentage point
should therefore not be over-interpreted.  } While independent historical replay is
still closest in marginal tail mass, the Tail Diffusion is a fully
parametric generator.

\section{Proof of The Conditional No Arb Test}
\label{sec:proof-cond-no-arb}

\begin{proof}[Proof of theorem~\ref{th:cond-no-arb}]
Because the objective is additive over $t$ and $\mathcal A$ contains all
bounded adapted actions, $a_t^*$ minimizes the corresponding conditional
objective pointwise.  For a fixed state define
\begin{equation}
  \ell_t(u)
  :=\frac12u'u+\log\E_t^\Q[\exp(-u'R_t)].
  \label{eq:conditional-action-log-objective}
\end{equation}
Let $\widetilde\Q_{t,u}$ denote the conditional law obtained by tilting
$\Q(\,\cdot\mid\mathcal F_t)$ proportionally to $\exp(-u'R_t)$.  Then
\begin{equation}
  \nabla\ell_t(u)
  =u-\E_t^{\widetilde\Q_{t,u}}[R_t],
  \qquad
  \nabla^2\ell_t(u)
  =I+\operatorname{Cov}_t^{\widetilde\Q_{t,u}}(R_t)
  \succeq I.
  \label{eq:conditional-action-derivatives}
\end{equation}
Thus $\ell_t$ is strictly convex, its minimizer is unique, and
$\nabla\ell_t(0)=-\mu_t$ for
$\mu_t:=\E_t^\Q[R_t]$.  Strong monotonicity of its gradient gives
\begin{equation}
  {a_t^*}'\mu_t
  =\left(\nabla\ell_t(a_t^*)-\nabla\ell_t(0)\right)'a_t^*
  \geq \lVert a_t^*\rVert^2,
  \label{eq:conditional-action-lower-bound}
\end{equation}
where $\nabla\ell_t(a_t^*)=0$.  Consequently,
\begin{equation}
  \E^\Q[{a_t^*}'R_t]
  =\E^\Q[{a_t^*}'\mu_t]
  \geq\E^\Q[\lVert a_t^*\rVert^2].
  \label{eq:captured-return-lower-bound}
\end{equation}
If $\mu_t=0$, strict convexity makes $a_t^*=0$, and the captured return is
zero.  Conversely, if the captured return is zero, the last display implies
$a_t^*=0$ almost surely; its first-order condition then gives $\mu_t=0$
almost surely.
\end{proof}

\section{Risk-neutral optimizer convergence}
\label{app:rn-q2-convergence}

\begin{figure}[H]
  \centering
  \includegraphics[width=0.82\textwidth]{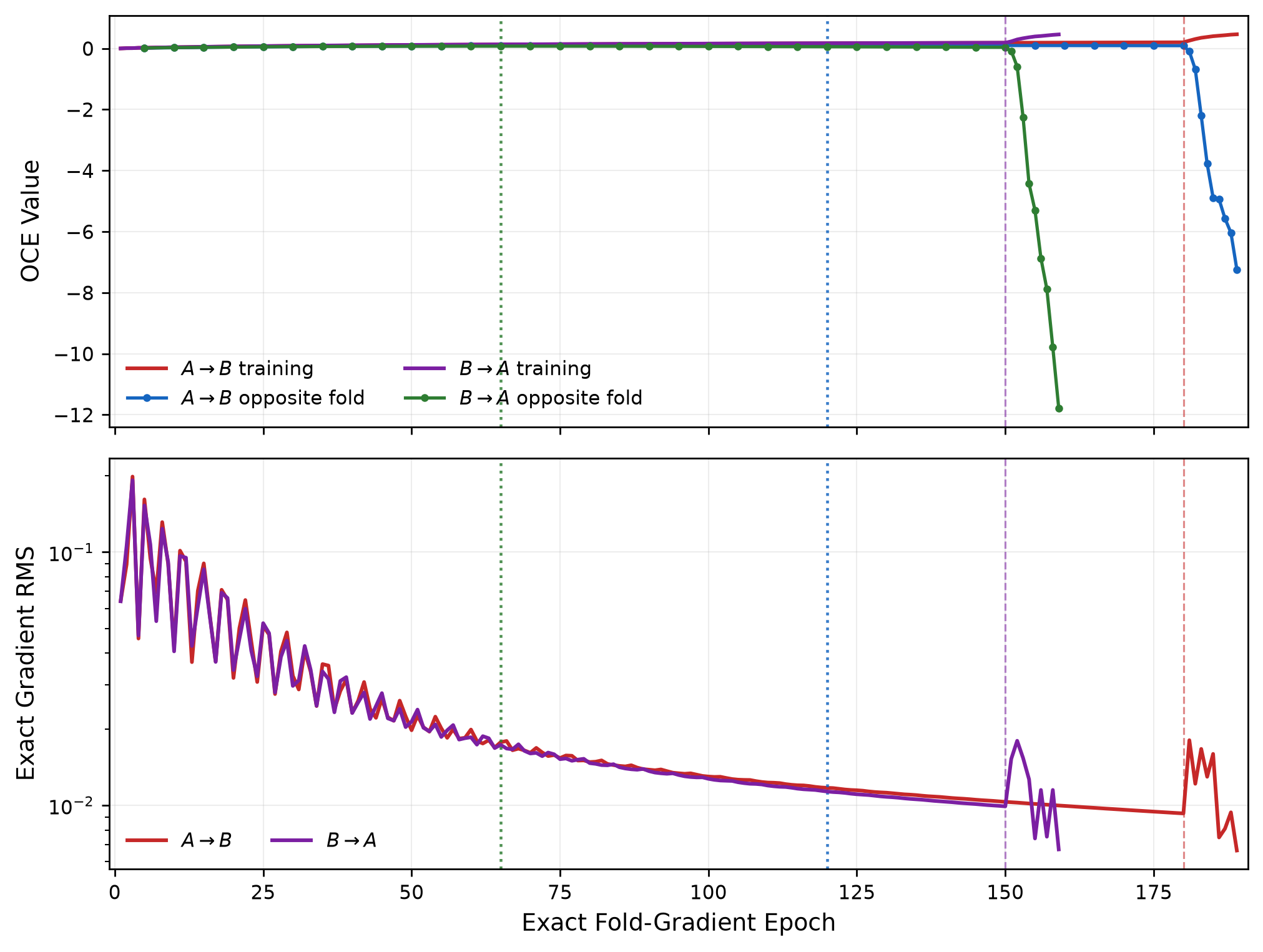}
  \caption{Full-population optimization of~\eqref{eq:rn-oce-q2-fit}.  The top
  panel shows the OCE objective and the bottom panel its exact parameter-gradient
  RMS against cumulative full-population objective/gradient evaluations.  The
  vertical dashed line marks the switch from Adam to L-BFGS; the horizontal
  dotted line is the terminal stopping target.}
  \label{fig:rn-q2-convergence}
\end{figure}

\section{Data pipeline}\label{app:data-pipeline}

In the interest of full transparency we here discuss main elements of the data pipeline used
in our experiments. \\

The market-data pipeline constructs a standardized daily time series of SPX option surfaces. Raw option quotes are cleaned, normalized, filtered, checked against an arbitrage-free reference surface, and finally collected into irregular quote books suitable for direct model training.

The pipeline processes SPX option data between January 2nd, 2020 and August 29th, 2025. The latter
is the latest day available at the time of writing. Only valid full-day NYSE trading days are considered.

\subsection{Pipeline Overview}

The complete transformation can be summarized as
\[
\begin{aligned}
\text{raw SPX option quotes}
&\longrightarrow \text{daily cleaning and enrichment} \\
&\longrightarrow \text{arbitrage-free quote surface using linear SANOS} \\
&\longrightarrow \text{liquidity and moneyness filters} \\
&\longrightarrow \text{shared 200-strike SANOS model geometry} \\
&\longrightarrow \text{direct native and affine-200 quote fits}.
\end{aligned}
\]
The key insight when working with Option Metrics IvyDB data is that the forward curves
from the \texttt{Forward\_Price} table do not line up with the option data themselves. We therefore
compute our own forward from weighted put-call parities. We also convert the tightest options into call prices. Such data still has arbitrage, hence we use a linear SANOS pre-processing step to obtain fitted prices within bid/ask which adhere to basic linear no-arbitrage conditions (the absence of which would indicate actionable arbitrage). We drop any strikes where such a fit
cannot be achieved, assuming that fit issues are due to data alignment challenges rather than
genuine trading opportunities.

\subsection{Raw Daily Market Data}

For every eligible date \(d\), the underlying data contain quantities such as
\begin{itemize}
    \item the SPX spot price \(S_d\) and previous spot price \(S_{d-1}\);
    \item option strikes and expiration dates; call and put bid and ask quotes; volume and open interest; AM/PM settlement indicators;
    \item forward and discount-curve information (with forwards being re-calculated below).
\end{itemize}

The default raw-data cleaning removes
\begin{itemize}
    \item same-day expirations;
    \item options with bids below \(0.011\) (the minimum tick size is 0.05\$ for series trading below 3\$);
    \item observations with missing or invalid quotes; observations with an invalid settlement flag; irregular expirations; dates or ticker-date combinations known to contain bad data (e.g.~2025-01-08 which was a holiday).
\end{itemize}

\subsection{Enrichment and Expiry Filtering}

The cleaned quotes are enriched with
\begin{itemize}
    \item business days and year fractions to expiry where AM options are counted as 20\% of
    the day;
    \item implied forward prices computed off put-call parity per expiry for all strikes
          for which both call and put were traded and have valid bid/ask prices and sizes, weighted by put-call implied forward bid-ask spread.\footnote{
            We compute
            $\mathrm{fwd}^\mathrm{ask}(K):=+\mathrm{Call}^\mathrm{ask}-\mathrm{Put}^\mathrm{bid}-\mathrm{df} K$ and
            $\mathrm{fwd}^\mathrm{bid}(K):=+\mathrm{Call}^\mathrm{bid}-\mathrm{Put}^\mathrm{ask}-\mathrm{df} K$ using the discount factor sourced from IvyDB. 
            Then we
            define $\mathrm{fwd}^\mathrm{mid}(K) := 0.5\,(\mathrm{fwd}^\mathrm{ask}(K) + \mathrm{fwd}^\mathrm{bid}(K))$ and $\gamma^\mathrm{mid}(K) := 0.5\,(\mathrm{fwd}^\mathrm{ask}(K) - \mathrm{fwd}^\mathrm{bid}(K))$ and
            $$
                \mathrm{fwd} :=\frac{ \sum_K \gamma(K)^{-1} \mathrm{fwd}^\mathrm{mid}(K) }
                 { \sum_K \gamma(K)^{-1} }
            $$
          }
    \item using the new forward, define `market' mid call prices between the actual call
    prices and synthetic call prices from the puts;
    \item compute implied volatilities; forward- and discount-adjusted prices, strikes, and Greeks.
\end{itemize}
An expiration is retained only if it satisfies the default liquidity and quality requirements. In particular,
\begin{itemize}
    \item at least \(10\) strikes must have either a traded call or put;
    \item at least \(4\) strikes must have both a traded call and put;
    \item the discounted forward normalized spread must be at least \(10^{-6}\);
    \item implied volatility must lie in
    \[
        0.01 \leq \sigma^{\mathrm{impl}} \leq 5.
    \]
\end{itemize}

\subsection{Arbitrage-Free Preprocessing}

Before the final option filter is applied, an arbitrage-free representation of the daily market is constructed using~\citet{SANOS2026}. The default configuration uses surface mode and performs the following operations:
\begin{enumerate}
    \item Extend the strike grid to improve boundary treatment.
    \item Fit a linear (DLV) no-arbitrage option-price curve separately for each expiry.
    \item Drop expiries for which a valid fit cannot be obtained.\footnote{
        Existence of such a fit is equivalent to the market not having actionable arbitrage
        using bid/asks as quoted.    
    }
    \item Remove locally redundant strikes associated with negligible fitted density.
    \item Fit a surface across strike and expiry to control both butterfly and calendar arbitrage.
\end{enumerate}

At least five strikes per expiry are required. The default minimum fitted density is \(10^{-4}\), and the minimum admissible ATM volatility is \(0.05\).

\subsection{Option Filtering}

The empirical analysis retains observations satisfying
\[
    \operatorname{BDTE} \leq 70,
\]
where \(\operatorname{BDTE}\) denotes business days to expiry.

The normalized-moneyness filter is
\[
    -3 \leq z_{\mathrm{ATM}} \leq 2,
\]
where \(z_{\mathrm{ATM}}\) is normalized moneyness computed using the
arbitrage-free fitted ATM volatility for the corresponding expiry.

Additional requirements are
\[
\begin{aligned}
    \mathrm{VegaSqrtVar} &\geq 10^{-3}, \\
    \log(\mathrm{volume}) &\geq 1, \\
    \log(\mathrm{open\ interest}) &\geq 1.
\end{aligned}
\]

Here, `$\mathrm{VegaSqrtVar}$' refers to Vega divided by sqrt of expiry. It is therefore
the sensitivity to total volatility of the option.

Each retained expiry must contain at least \(10\) options.

It should be noted that this filter removes many options post 1Y for the early years of the training period due to the minimum volume
traded filter. We suspect that this information in IvyDB is not reliable in the past.
Future tests might want to remove the volume and/or open interest filters. (It is an option
in our data pipeline parametrization.)

\subsection{Spot Normalization}

Price-like quantities are normalized by the current spot \(S_d\) to obtain
variables of scale around~$1$. For example,
\[
    \widetilde{K}_d^i
    =
    \frac{K_d^i}{S_d},
    \qquad
    \widetilde{C}_d^i
    =
    \frac{C_d^i}{S_d},
    \qquad
    \widetilde{F}_d^i
    =
    \frac{F_d^i}{S_d}.
\]
The original spot is retained as
$
    \mathrm{spot\_ref}_d = S_d.
$
(The normalized log-spot level is passed on as $h^0$ later.)

\subsection{Arbitrage-free preprocessing and model geometry}

The preceding steps produce tradable normalized calls and puts with their
observed bid--ask intervals.  A per-date arbitrage-free SANOS fit supplies a
smooth reference surface, an ATM-volatility curve, and an availability check
for the retained quotes.  The completed Direct-200 experiments do
\emph{not} resample this information onto a secondary reference grid.
Instead, every eligible original quote enters the loss at its observed strike
and expiry.

The decoder uses the 18 model expiries stated in Section~\ref{sec:data}:
1D--5D followed by 10D, 15D, and five-session increments through 70D.  On
each date it builds one shared support with $N=200$ inner strikes. Half of the normalized-moneyness
candidates cover the first model expiry and half cover the final model expiry
over $z\in(-3,2)$.  The candidates are merged and spaced, and the two boundary
strikes are
\[
  \hat K_t^0=\tfrac12\hat K_t^1,
  \qquad
  \hat K_t^{N+1}=\tfrac32\hat K_t^N.
\]
The same support is used at every model expiry on that date; only the density
and DLV coordinates carry an expiry superscript.

\subsection{Fitting ML-SANOS}

The native control first fits one unrestricted vector
$u_t\in\R^{M(N+1)+2}$ to each date's irregular quote book.  The completed
affine model then uses the map
\[
  h_t\in\R^{200}
  \longmapsto
  u_t=b+Bh_t\in\R^{M(N+1)+2}
\]
from~\eqref{eq:affine-decoder}.  It is initialized from the native controls
and first fine-tuned across all 891 dates using only the market objective.
The stable AR overlay then jointly refines this affine model and its one-week
transition.  Prices are evaluated directly at every observed quote; the
training weights and inside-spread objective are those in
\eqref{eq:production-quote-loss}.  Appendix~\ref{app:affine-training}
describes the complete optimization sequence.

\clearpage
\section{Option-surface Literature Classification}
\label{app:literature-tables}
Tables~\ref{tab:surface-interpolation-literature},
\ref{tab:nonparametric-surface-interpolation-literature},
\ref{tab:surface-representation-literature},
\ref{tab:inverse-calibration-literature}, and
\ref{tab:surface-dynamics-literature} separate four questions which are often
conflated: constructing an option or implied volatility surface, compressing or generating a
market state, solving an underlying-model inverse problem, and learning a
surface's evolution through time.  We use common classification
symbols throughout the tables:
\begin{itemize}[leftmargin=*,nosep]
  \item \textbf{S$\infty$}: a theorem- or architecture-level guarantee on the
        full continuous surface.
  \item \textbf{S$\Omega$}: the same guarantee only on a stated bounded fitting
        domain.
  \item \textbf{SG}: a strict guarantee only on the modelled finite grid.
  \item \textbf{A}: approximate, using penalties, reweighting, rejection, or
        empirical violation rates.
  \item \textbf{N}: no static-arbitrage guarantee.
  \item \textbf{U}: arbitrage freedom is claimed, but the available construction
        does not suffice to verify its continuum scope.
\end{itemize}

\newcommand{\auditnew}[1]{#1}
\newcolumntype{L}[1]{>{\raggedright\arraybackslash}p{#1}}

\begin{landscape}
\begin{table}[p]
\centering
\fontsize{7}{7.8}\selectfont
\setlength{\tabcolsep}{3pt}
\renewcommand{\arraystretch}{1.04}
\caption{Selected parametric implied-volatility and option-price interpolation
methods.  ``Parametric'' means that the surface is restricted to a prescribed
finite-dimensional functional family; it does not describe the calibration
algorithm. Constraints imposed during fitting are distinguished from post-fit repair.}
\label{tab:surface-interpolation-literature}
\begin{tabular}{@{}L{4.2cm}L{3.8cm}L{2.2cm}L{3.2cm}L{2.0cm}L{8.0cm}@{}}
\toprule
Method & Representation & Static arb. & Native inputs & Off-grid & Scope and qualification \\
\midrule
\auditnew{\citet{Gatheral2004SVI}}
  & \auditnew{SVI smile} & \auditnew{N by itself}
  & \auditnew{Arbitrary, one expiry} & \auditnew{$\infty$ per expiry}
  & \auditnew{Introduces raw SVI and its wing asymptotics, but does not supply the later cross-expiry SSVI construction or a complete admissible parameter domain.} \\
\citet{Hafner2004}
  & Polynomial IV surface & N & Variable daily DAX transactions & $\infty$ on bounded domain
  & Weighted least-squares fit followed by finite-grid screening: dates violating hedge, bull-spread, butterfly, or calendar tests on a 200-point grid are discarded.  The fitted family itself is unconstrained. \\
\citet{Bloch2013MixVar}
  & Mixture family & U & Arbitrary & $\infty$
  & Claimed arbitrage-free by construction; the continuum conditions were not verified. \\
\citet{GatheralJacquier2014}
  & SVI/SSVI & S$\infty$ & Arbitrary & $\infty$
  & Strict for admissible parameters; a central low-dimensional benchmark. \\
\citet{GuoJacquierMartiniNeufcourt2016}
  & Generalized SVI & S$\infty$ & Arbitrary & $\infty$
  & Extends the strict SVI/SSVI constructions. \\
\auditnew{\citet{CorbettaEtAl2019}}
  & \auditnew{SSVI slices} & \auditnew{S$\infty$}
  & \auditnew{Arbitrary maturities and strikes} & \auditnew{$\infty$}
  & \auditnew{Calibrates butterfly- and calendar-consistent slices, then proves that its parameter interpolation and extrapolation preserve those conditions over the continuous surface.} \\
\auditnew{\citet{HendriksMartini2019}}
  & \auditnew{eSSVI} & \auditnew{S$\infty$ (conditional)}
  & \auditnew{Arbitrary} & \auditnew{$\infty$}
  & \auditnew{Explicit butterfly and calendar conditions; strictness requires admissible parameter curves.} \\
\auditnew{\citet{MartiniMingone2022}}
  & \auditnew{SVI smile} & \auditnew{S$\infty$ per expiry}
  & \auditnew{Arbitrary, one expiry} & \auditnew{$\infty$ per expiry}
  & \auditnew{Characterizes the no-butterfly domain of raw SVI and calibrates directly on it; calendar consistency across independently fitted expiries is not automatic.} \\
\auditnew{\citet{Mingone2022}}
  & \auditnew{eSSVI surface} & \auditnew{S$\infty$}
  & \auditnew{Arbitrary} & \auditnew{$\infty$}
  & \auditnew{A global eSSVI parametrization whose admissible parameter domain enforces butterfly and calendar-spread conditions.} \\
\citet{Roos2025}
  & Discrete-time local-vol family & U & Arbitrary coterminal set & $\infty$
  & The abstract claims analytic strike interpolation and expiry extension; the continuum inequalities were not verified. \\
\bottomrule
\end{tabular}
\end{table}
\end{landscape}

\begin{landscape}
\fontsize{7}{7.7}\selectfont
\setlength{\tabcolsep}{2.5pt}
\renewcommand{\arraystretch}{1.04}
\begin{longtable}{@{}L{4.2cm}L{4.0cm}L{2.2cm}L{3.4cm}L{2.0cm}L{7.6cm}@{}}
\caption{Selected non-parametric implied-volatility and option-price
interpolation methods.  Soft-penalty methods are approximate even when reported
violations are negligible.}
\label{tab:nonparametric-surface-interpolation-literature}\\
\toprule
Method & Representation & Static arb. & Native inputs & Off-grid & Scope and qualification \\
\midrule
\endfirsthead
\multicolumn{6}{c}{\tablename\ \thetable\ -- continued}\\
\toprule
Method & Representation & Static arb. & Native inputs & Off-grid & Scope and qualification \\
\midrule
\endhead
\midrule
\multicolumn{6}{r}{Continued on next page}\\
\endfoot
\bottomrule
\endlastfoot
\auditnew{\citet{Kahale2004}}
  & \auditnew{Piecewise analytic call/IV interpolation}
  & \auditnew{S$\Omega$ (conditional)} & \auditnew{Arbitrage-free discrete quotes}
  & \auditnew{$\infty$ for $T\leq T_{\max}$}
  & \auditnew{Convex-envelope construction on the bounded maturity domain; no extension beyond the last input expiry.} \\
\citet{fengler2009}
  & Constrained spline & S$\Omega$ per expiry; SG calendar & Arbitrary & $\infty$ on fitted strike domain
  & Strike constraints hold over each fitted spline; calendar constraints only at knots and input maturities. \\
\citet{AndreasenHuge}
  & One-step implicit FD; basis for DLV & SG numerically; S$\infty$ for exact local-vol model & Arbitrary quotes & $\infty$
  & Sequential local-vol calibration; the finite-difference scheme preserves monotonicity and convexity on its grid. \\
\citet{Orosi2015}
  & Thin-plate regression splines/QP & U & Arbitrary & $\infty$
  & Quadratic-program constraints; their continuum scope was not verified. \\
\citet{AckererTagasovskaVatter2020}
  & Neural network & A & Irregular daily panel & $\infty$
  & Static conditions enter through soft penalties; violation rates depend on the penalty weight. \\
\auditnew{\citet{BenderThiel2020}}
  & \auditnew{Reference-model distortion of call prices}
  & \auditnew{S$\infty$ (conditional)} & \auditnew{Arbitrage-free discrete quotes}
  & \auditnew{$\infty$}
  & \auditnew{Smooth strike interpolation and arbitrage-free maturity interpolation.} \\
\citet{ZhangLiZhang2021}
  & PCA/VAE/samples plus neural surface & A & $14\times11=154$ & $\infty$
  & Calendar, butterfly, and wing conditions enter as penalties; occasional spikes remain. \\
\citet{ChataignerEtAl2022}
  & Constrained GP or penalized NN & GP: S$\Omega$; NN: A & Arbitrary & $\infty$
  & Nodal inequalities imply monotonicity and convexity throughout the GP's bounded piecewise-multilinear domain.  The NN and fitted SSVI alternatives use penalties. \\
\auditnew{\citet{Deschatres2024}}
  & \auditnew{Variance-space cubic/B-spline QP} & \auditnew{A}
  & \auditnew{Arbitrary; 14,500 quotes in the SPX example} & \auditnew{$\infty$}
  & \auditnew{Calendar constraints are linear, but butterfly constraints are linearized and the QP is iterated.  Under the present convention this is approximate rather than a single exact admissible fit.} \\
\citet{YangChenShuHospedales2025}
  & Hypernetwork plus compact neural surface & A & 9 observations & $\infty$
  & Reconstructs index-option IV surfaces from sparse option quotes.  Calendar and density conditions are finite-grid auxiliary losses; the paper explicitly states that they provide no hard guarantee. \\
\auditnew{\citet{GononJacquierWiedemann2025}}
  & \auditnew{Graph neural operator} & \auditnew{A}
  & \auditnew{Irregular intraday panels} & \auditnew{$\infty$ on its output domain}
  & \auditnew{Calendar and butterfly conditions are finite-difference penalty losses evaluated at sampled coordinates; the paper describes the output as virtually arbitrage-free, not structurally invariant.} \\
\citet{Schlenkrich2025}
  & Piecewise shifted-Black/Bachelier model & S$\infty$ per expiry; calendar N & Arbitrary & $\infty$ per expiry
  & Whole-strike construction for one terminal distribution; calendar consistency is not addressed. \\
\citet{QinYangCheFeng2026}
  & Piecewise-quadratic calls/RN marginals & S$\infty$ in strike at quoted expiries & Arbitrage-free discrete quotes & $\infty$ in strike only
  & Explicit sequential completion of risk-neutral marginals; no interpolation between expiries. \\
\citet{SANOS2026}
  & Smooth non-parametric DLV & S$\infty$ & Arbitrary; about 1,000 in example & $\infty$
  & Bid--ask-aware smooth surface with positivity as the essential parameter constraint. \\
\end{longtable}
\end{landscape}

\begin{landscape}
\begin{table}[p]
\centering
\fontsize{7}{7.8}\selectfont
\setlength{\tabcolsep}{2.5pt}
\renewcommand{\arraystretch}{1.06}
\caption{PCA, autoencoder, and other latent representations of option surfaces.
``Off-grid'' counts options not represented by native model coordinates;
external interpolation is not credited to the representation.
  An off-grid count of
$\infty$ means that the representation can be evaluated on a continuum, subject
to its stated domain and tail assumptions.
}
\label{tab:surface-representation-literature}
\begin{tabular}{@{}L{4.3cm}L{4.2cm}L{3.6cm}L{2.7cm}L{3.0cm}L{6.0cm}@{}}
\toprule
Method & Representation & Native option coordinates & Off-grid & Latent/static arb. & Qualification \\
\midrule
\citet{SkiadopoulosHodgesClewlow2000}
  & Direct-IV PCA & Not reported; three maturity groups & 0 & N & PCA factors do not preserve static inequalities. \\
\citet{ContVol}
  & Functional/direct-IV PCA & Not reported & Any number within the fitted domain & N & Functional interpolation does not make the PCA cone invariant. \\
\auditnew{\citet{FenglerHardleVilla2003}}
  & \auditnew{Common PCA of direct IV changes} & \auditnew{$6\times6=36$: moneyness 0.85--1.10 at 1, 2, 3, 6, 9, and 12 months (DAX)}
  & \auditnew{0} & \auditnew{N}
  & \auditnew{Common factors summarize smile dynamics but do not define an invariant no-arbitrage state space.} \\
\citet{Hafner2004}
  & Four parametric IV factors: level, slope, curvature, and term slope & Variable daily DAX transactions; 200-point diagnostic grid & $\infty$ on its bounded domain & N & The daily regression
   family is not constrained.  Dates failing finite-grid hedge, bull-spread, butterfly, or calendar tests are discarded ex post. \\
\citet{PanigirtzoglouSkiadopoulos2004}
  & Separate implied-distribution PCA by expiry & $3\times21=63$: 21 deltas at 1, 3, and 6 months & 0 & PDF: S$\infty$ per expiry; CDF: N; joint surface: N & The PDF construction is strict per expiry; neither construction enforces calendar consistency. \\
\citet{DellaportasMijatovic2014}
  & Parametric SABR encoding & SABR triplet for one expiry & Any strike in fitted interval & U & SABR may violate arbitrage conditions in the extreme wings. \\
\citet{wiese2019simulate}
  & DLV, PCA, and GAN & $8\times4=32$ & 0 & SG & Finite DLV lattice. \\
\auditnew{\citet{AvellanedaEtAl2020}}
  & \auditnew{PCA of normalized IV-surface deformations} & \auditnew{$8\times7=56$ per name: 8 maturities and 7 deltas}
  & \auditnew{0} & \auditnew{N}
  & \auditnew{A data-driven factor representation; no static-arbitrage invariant decoder is imposed.} \\
\auditnew{\citet{BergeronEtAl2021}}
  & \auditnew{VAE representation of IV surfaces} & \auditnew{$8\times5=40$: 8 maturities and 5 deltas (FX)}
  & \auditnew{Any (delta, maturity) in the domain via the pointwise decoder} & \auditnew{N}
  & \auditnew{A hands-off latent representation and generator, without a hard static-arbitrage architecture.} \\
\citet{wiese2021multiassetspotoptionmarket}
  & DLV, constrained autoencoder, and flow & $9\times4=36$ per asset & 0 & SG & Decoder is constrained on the finite lattice. \\
\citet{ZhangLiZhang2021}
  & PCA/VAE/samples plus neural reconstruction & 154 & $\infty$ & A & Reconstruction uses static-arbitrage penalties. \\
\citet{AndresBoumezouedJourdain2023}
  & Four-parameter SSVI state & 384 (Euro Stoxx 50); 1,944 (SPX, before filtering) & $\infty$ & S$\infty$ & Admissible SSVI parameter conditions hold for each state. \\
\citet{Cohen02012023}
  & Affine factors of normalized call prices & 46 options on an irregular rolling $(\tau,m)$ lattice & 0 & SG & Linear no-static-arbitrage inequalities define a factor polytope. \\
\citet{ContVuletic2023}
  & Direct-IV PCA/factors plus path reweighting & Fixed IV grid & 0 & Raw PCA: N; output: A & Arbitrage scenarios are downweighted, not excluded structurally. \\
\citet{ChoudharyJaimungalBergeron2024}
  & Functional PCA and neural SDE & $17\times11=187$ per asset & Any number within the fitted domain & N & Static-arbitrage violations are measured only ex post. \\
\citet{VuleticCont2024}
  & Direct-IV GAN plus penalties/reweighting & $10\times8=80$ & 0 & A & Approximate output correction. \\
\citet{JinAgarwal2025}
  & Direct-IV conditional diffusion & $9\times9=81$ & 0 & A & Penalty-based. \\
\citet{GolubitskyDavisonCaraman2026}
  & VAE latent representation & $2\times11\times17=374$ & 0 & A & Static constraints are approximate. \\
\citet{BrooksEtAl2026}
  & Arbitrage-penalized VAE and latent flow & $32\times16=512$ & 0 & A; 90.8\% pass all tests & Empirical pass rate, not a guarantee. \\
\citet{HaoJi2026}
  & Conditional diffusion plus SAAM refinement & $15\times17=255$ & Arbitrary target coordinates & A & SAAM uses residual regularization and reduces measured violations on a finite diagnostic grid; it is not an admissible parameterization. \\
\citet{HanEtAl2026}
  & Joint return/direct-IV adaptive sequential diffusion & $11\times9=99$ IVs & 0 & A & Fine-tuning penalizes calendar-, call-, and butterfly-spread violations; reported violations are nearly zero, not structurally excluded. \\
\citet{YangWangZhangLi2026}
  & Joint stock-return/option-IV diffusion across equities & $11\times9=99$ option IVs per stock & 0 & A & Transfers dynamics to unseen stocks on the same fixed grid. \\
This paper
  & Smooth SANOS plus affine latent state & $18\times200=3{,}600$ prices; 3,620 controls & $\infty$ & S$\infty$ & The decoded continuum remains in the SANOS admissible family. \\
\bottomrule
\end{tabular}
\end{table}
\end{landscape}

\begin{landscape}
\begin{table}[p]
\centering
\fontsize{7}{7.8}\selectfont
\setlength{\tabcolsep}{2pt}
\renewcommand{\arraystretch}{1.08}
\caption{Machine-learning methods for underlying-model inverse problems.
These methods infer or optimize parameters of a pre-specified pricing model;
they do not make the observed option market itself the dynamic state.  Static
arbitrage properties are inherited from the pricing model, subject to decoded
parameters remaining admissible.}
\label{tab:inverse-calibration-literature}
\begin{tabular}{@{}L{4.5cm}L{5.0cm}L{5.2cm}L{4.2cm}L{5.2cm}@{}}
\toprule
Method & Underlying pricing family & Role of ML & Online inverse step & Why separate from a market-state model \\
\midrule
\citet{Hernandez2016}
  & Hull--White/G2++ examples & Direct inverse map trained on model-generated examples & One network evaluation & Maps quotes to parameters of a prescribed model. \\
\auditnew{\citet{DimitroffRoederFries2018}}
  & \auditnew{Heston}
  & \auditnew{Convolutional network maps volatility grids to model parameters}
  & \auditnew{One network evaluation}
  & \auditnew{Amortized inversion into a prescribed pricing family.} \\
\citet{BayerStemper2018}
  & Rough Bergomi and related rough-volatility models & Learns the forward parameter-to-IV map & Classical Levenberg--Marquardt optimization through the surrogate & The NN accelerates a conventional model calibration. \\
\citet{HorvathMuguruzaTomas2019,BayerEtAl2019DeepCalibration}
  & Broad class including rough-volatility models & Learns a fast forward pricing/IV map & Classical numerical or Bayesian inversion of the surrogate & Explicitly retains the two-step pricing-model-plus-calibration architecture. \\
\auditnew{\citet{LiuEtAl2019Calibration}}
  & \auditnew{Heston and Bates}
  & \auditnew{Learns a forward parameter-to-observable map}
  & \auditnew{Differential evolution through the surrogate}
  & \auditnew{The learned map accelerates conventional parameter calibration.} \\
\citet{Itkin2019}
  & Black--Scholes examples & Compares direct and surrogate calibration & Direct map or numerical solver & Studies smooth Greeks, soft arbitrage penalties, and inverse-map stability. \\
\citet{RoederDimitroff2020}
  & Volatility-model examples used in deep-calibration benchmarks & Compares a direct IV-to-parameter map with the two-step forward surrogate & None for the direct version; solver for the indirect version & Both outputs are parameters of an exogenously specified pricing model. \\
\auditnew{\citet{CuchieroKhosrawiTeichmann2020}}
  & \auditnew{Local stochastic-volatility models}
  & \auditnew{Neural leverage function}
  & \auditnew{Monte Carlo SGD; adversarial optimization only for robust bid--ask calibration}
  & \auditnew{The generator remains an underlying-first LSV model.} \\
\citet{RosenbaumZhang2021}
  & Quadratic rough Heston & Neural pricing/calibration engine based on a multifactor approximation & Optimization against the learned engine & The calibrated state is a rough-volatility parameter vector. \\
\citet{vae_ning2023arbitrage}
  & Time-inhomogeneous CTMC; additive Gaussian-mixture and CGMY/KoBoL L\'evy models & VAE/PCA learns the cross-date distribution of already calibrated parameters & Regularized inverse calibration is done first, date by date; CTMC maturities are fitted sequentially & The VAE generates model parameters, and prices are recovered through the model's characteristic function. \\
\citet{SridiBilokon2023}
  & Heston & Neural pricing surrogate & Differential evolution and Nelder--Mead & Heston supplies the risk-neutral structure. \\
\citet{ZhangAmiciMorandotti2024}
  & Heston & Differential network learns prices and sensitivities & Gradient calibration through the learned pricer & Heston supplies the risk-neutral structure. \\
\bottomrule
\end{tabular}
\end{table}
\end{landscape}

\begin{landscape}
\fontsize{7}{7.8}\selectfont
\setlength{\tabcolsep}{2.2pt}
\renewcommand{\arraystretch}{1.06}
\begin{longtable}{@{}L{3.7cm}L{3.4cm}L{2.8cm}L{2.2cm}L{4.1cm}L{1.5cm}L{5.3cm}@{}}
\caption{Option-surface dynamics under the historical measure $\P$ and direct
market-model dynamics under $\Q$.  The $\Q$ column excludes underlying-first
pricing families such as Heston, local volatility, and Bates.  ``Multi-step
$\P$ path'' records whether the cited work actually reports a simulated or
forecast path spanning more than one transition, rather than whether its model
could in principle be iterated.  ``Implemented'' refers to the stated dynamic
construction, not merely static preprocessing.}
\label{tab:surface-dynamics-literature}\\
\toprule
Method & Dynamic state/model & Static arb. under $\P$ & Multi-step $\P$ path & Direct market-model dynamics under $\Q$ & Implemented & Qualification \\
\midrule
\endfirsthead
\multicolumn{7}{c}{\tablename\ \thetable\ -- continued}\\
\toprule
Method & Dynamic state/model & Static arb. under $\P$ & Multi-step $\P$ path & Direct market-model dynamics under $\Q$ & Implemented & Qualification \\
\midrule
\endhead
\midrule
\multicolumn{7}{r}{Continued on next page}\\
\endfoot
\bottomrule
\endlastfoot
\auditnew{\citet{Schonbucher1999}}
  & \auditnew{One fixed-strike term structure of implied volatilities}
  & \auditnew{---} & \auditnew{---} & \auditnew{Exact drift restrictions across maturities for one strike}
  & \auditnew{MC implementation strategy only}
  & \auditnew{The setup initially permits a generic finite option set, but the developed discrete and continuous constructions vary maturity at one strike.  Independent dynamics across strikes at the same maturity are explicitly left for future research; this is not a full-surface model.} \\
\citet{SkiadopoulosHodgesClewlow2000}
  & Direct-IV PCA factors & N & No & --- & Yes & No static invariant state space. \\
\citet{ContVol}
  & Functional/direct-IV PCA factors & N & No & --- ($\P$ only) & Empirical PCA only & No constructed admissible $\Q$ dynamics. \\
\auditnew{\citet{ContDaFonseca2002Dynamics}}
  & \auditnew{Functional PCA factors of direct IV surfaces} & \auditnew{N}
  & \auditnew{No} & \auditnew{---} & \auditnew{Yes ($\P$)}
  & \auditnew{Empirical factor dynamics; distinct from the three-author stochastic-surface paper and not a constructed risk-neutral model.} \\
\auditnew{\citet{FenglerHardleVilla2003}}
  & \auditnew{Common PCA factors of direct IV changes} & \auditnew{N}
  & \auditnew{No} & \auditnew{---} & \auditnew{Yes ($\P$)}
  & \auditnew{Empirical CPCA of observed daily surfaces.  The factors are proposed as a starting point for VaR Monte Carlo, but the paper does not report synthetic multi-step IV paths.  There is no invariant static-arbitrage state space.} \\
\citet{Hafner2004}
  & Four IV factors with correlated OU dynamics & N; screened sample only & No (1-, 5-, and 10-day forecasts) & A/local & Yes & The fitted OU model violates the global drift condition; the $\Q$ construction is local and approximate. \\
\citet{PanigirtzoglouSkiadopoulos2004}
  & Separate distribution factors by constant maturity & PDF: S$\infty$ per expiry; CDF: N & Yes (14 days) & N for a joint surface & Yes & Neither construction enforces calendar consistency. \\
\auditnew{\citet{FenglerHardleMammen2007}}
  & \auditnew{Semiparametric IV-factor surface} & \auditnew{N}
  & \auditnew{No} & \auditnew{---} & \auditnew{Yes ($\P$)}
  & \auditnew{Estimates and forecasts historical surface factors without a hard static-arbitrage invariant.} \\
\citet{SchweizerWissel2008}
  & Price level and local implied-volatility field & S$\infty$, theoretical & --- & S$\infty$: solvability theorem and explicit admissible coefficient examples & No & One fixed expiry and all strikes; not a finite option lattice or full maturity surface. \\
\auditnew{\citet{SchweizerWisselTerm2008}}
  & \auditnew{Term structure for one fixed payoff} & \auditnew{S$\infty$, theoretical}
  & \auditnew{No} & \auditnew{Strict theoretical drift/existence results}
  & \auditnew{No}
  & \auditnew{Existence for fixed-strike calls and a power payoff; not a full-surface model.} \\
\auditnew{\citet{CarmonaNadtochiy2009}}
  & \auditnew{Local-volatility surface as option-market codebook} & \auditnew{---}
  & \auditnew{---} & \auditnew{Drift and spot-consistency characterization}
  & \auditnew{Static empirical fit only}
  & \auditnew{Direct HJM-style codebook dynamics; general model existence is not proved.} \\
\auditnew{\citet{TangentLevy}}
  & \auditnew{Tangent L\'evy option-market codebook} & \auditnew{---}
  & \auditnew{---} & \auditnew{S$\infty$ under the consistency conditions}
  & \auditnew{Theory}
  & \auditnew{Constructs direct risk-neutral dynamics for all strikes and maturities from admissible tangent L\'evy characteristics.} \\
\citet{DellaportasMijatovic2014}
  & Single-expiry SABR parameter process & U & No (1--3-day point forecasts) & --- & Yes ($\P$) & No joint surface; extreme-wing arbitrage is not excluded. \\
\auditnew{\citet{KallsenKruhner2015}}
  & \auditnew{Time-inhomogeneous L\'evy option codebook} & \auditnew{---}
  & \auditnew{---} & \auditnew{S$\infty$: necessary/sufficient conditions and existence}
  & \auditnew{No numerical implementation}
  & \auditnew{Constructive existence and uniqueness from stated building blocks, but a theoretical HJM framework rather than a calibrated simulator.} \\
\auditnew{\citet{CarmonaMaNadtochiy2017}}
  & \auditnew{Finite-dimensional tangent L\'evy specifications} & \auditnew{---}
  & \auditnew{No} & \auditnew{S$\infty$ under model admissibility}
  & \auditnew{Yes}
  & \auditnew{Risk-neutral HJM drift with PCA-estimated volatility; eight-day $\Q$ paths.} \\
\pagebreak[4]
\citet{wiese2019simulate}
  & DLV PCA/GAN state & SG & Yes & --- & Yes & Finite-grid historical dynamics. \\
\auditnew{\citet{CaoChenHull2020}}
  & \auditnew{Neural regression of one-day IV changes} & \auditnew{N}
  & \auditnew{No} & \auditnew{---} & \auditnew{Yes ($\P$)}
  & \auditnew{Expected changes depend on return, delta/moneyness, maturity, and VIX; no arbitrage constraint.} \\
\citet{wiese2021multiassetspotoptionmarket}
  & Constrained DLV latent flow & SG & Yes & --- & Yes & Finite-grid historical dynamics. \\
\citet{ZhangLiZhang2021}
  & LSTM latent evolution and neural reconstruction & A & No (next-day only) & --- & Yes & Static-arbitrage penalties only. \\
\auditnew{\citet{removedrift}}
  & \auditnew{General $\P$-simulator plus an OCE/entropy path-density model}
  & \auditnew{Inherited from the input simulator}
  & \auditnew{Input-dependent}
  & \auditnew{Numerical near-martingale measure for a finite trading universe}
  & \auditnew{Yes}
  & \auditnew{Reweights an input simulator; with frictions, drift is targeted within transaction costs for selected instruments.} \\
\citet{MedvedevWang2022}
  & Direct-IV LSTM/ConvLSTM & N & No (multihorizon forecasts) & --- & Yes ($\P$) & No static-arbitrage constraint. \\
\citet{AndresBoumezouedJourdain2023}
  & Path-dependent SSVI factors & S$\infty$ for every simulated snapshot & Yes & No market-measure dynamics & Yes ($\P$) & Static arbitrage freedom does not imply dynamic no-arbitrage. \\
\citet{Cohen02012023}
  & Rolling normalized-call lattice, affine factors, and neural SDE & SG under $\P$: exact for model SDE & Yes & --- (not implemented) & Yes ($\P$ only) & The drift points inward and the normal diffusion vanishes at every face, so the continuous-time factor process remains in the finite-lattice no-static-arbitrage polytope $\P$-a.s.  The coordinates float with fixed $(\tau,m)$ rather than tracking fixed contracts.  The HJM expression is only a factor-selection criterion and residual diagnostic, not an implemented $\Q$ model. \\
\citet{ContVuletic2023}
  & Direct-IV factors plus path reweighting & A & Yes & --- & Yes & Reweighting reduces violations. \\
\citet{ChoudharyJaimungalBergeron2024}
  & Functional PCA and neural SDE & N & Yes & --- & Yes & Violations are measured only ex post. \\
\citet{VuleticCont2024}
  & Direct-IV GAN dynamics & A & No (next-day only) & --- & Yes & Penalties and reweighting. \\
\citet{JinAgarwal2025}
  & Conditional diffusion on IV grids & A & No (next-day only) & --- & Yes & Penalty-based. \\
\citet{BrooksEtAl2026}
  & Arbitrage-penalized VAE and latent flow matching & A & No & --- & Yes & Only 90.8\% of generated surfaces pass all reported tests. \\
\citet{HaoJi2026}
  & Conditional diffusion plus SAAM refinement & A & Yes (five one-minute steps) & --- & Yes & The daily experiment is next-day only; finite-grid violations remain. \\
\citet{HanEtAl2026}
  & Joint daily return and direct-IV grid via adaptive sequential diffusion & A & Yes (reported three-day path) & --- & Yes ($\P$) & Each step conditions on the updated generated history; hedging uses one-day conditional scenarios. \\
\citet{YangWangZhangLi2026}
  & Joint stock-return/option-IV diffusion across equities & A & No (next-day only) & --- & Yes & Jointly generates the next-day stock return and $11\times9$ option-IV increment; static option constraints are approximate. \\
\citet{FrancoisEtAl2026}
  & Five-factor IV surface jointly with SPX returns & A & No (five-day terminal scenarios) & --- & Yes ($\P$) & Rare empirical violations, not an invariant state space. \\
\citet{BucheggerGonon2026}
  & Latent diffusion for returns and IV surfaces & A & Yes (30-day trajectories) & --- & Yes ($\P$) & Penalty-based arbitrage awareness; violations remain. \\
This paper
  & SANOS, affine latent state, and AR(1) & S$\infty$ & Yes & Limited SG: 48 options plus spot & Yes & The implemented $\Q$ construction is finite-dimensional. \\
\end{longtable}
\end{landscape}

\end{document}